\documentclass[10pt,journal,letterpaper,compsoc]{IEEEtran}

\usepackage{cite}
\usepackage{url}
\usepackage[pdftex]{graphicx}
\usepackage[cmex10]{amsmath}
\usepackage{algorithm}
\usepackage{algorithmic}
\usepackage{array}
\usepackage[caption=false,font=footnotesize,labelfont=sf,textfont=sf]{subfig}
\usepackage{fixltx2e}
\usepackage{stfloats}
\usepackage{multirow}
\usepackage{color}
\usepackage{amssymb}
\usepackage[english]{babel}
\usepackage{amsthm}

\newtheorem{theorem}{Theorem}
\newtheorem{lemma}{Lemma}

\newtheorem{definition}{Definition}
\newtheorem{proposition}{Proposition}
\newtheorem*{game}{Game}

\ifodd 121

\newcommand{\com}[1]{\textbf{\color{red} (Comment: #1)}}
\newcommand{\comlin}[1]{\textbf{\color{magenta} (Lin Comment: #1)}}
\else

\newcommand{\com}[1]{}
\newcommand{\comlin}[1]{}
\fi
\def\K{\mathcal{K}}
\def\Ks{\K_{\textsc{s}}}

\def\Ka{\K_{\textsc{a}}}
\def\Ku{\K_{\textsc{u}}}
\def\M{\mathcal{M}}

\def\L{\textsc{l}}
\def\B{\textsc{b}}

\begin{document}

\title{Economic Analysis of Crowdsourced Wireless Community Networks}

\author{Qian~Ma,~\IEEEmembership{Student Member,~IEEE,}
		Lin~Gao,~\IEEEmembership{Member,~IEEE,}
        Ya-Feng~Liu,~\IEEEmembership{Member,~IEEE,}
        and~Jianwei~Huang,~\IEEEmembership{Fellow,~IEEE} 
\IEEEcompsocitemizethanks{
\IEEEcompsocthanksitem
Q. Ma and J. Huang are with Department of
Information Engineering, The Chinese University of Hong Kong. E-mail: \{mq012,\ jwhuang\}@ie.cuhk.edu.hk.
\IEEEcompsocthanksitem
L. Gao (Corresponding Author) is with School of Electronic and Information Engineering, Harbin Institute of Technology (Shenzhen), China. E-mail: 	
gaolin@hitsz.edu.cn.
\IEEEcompsocthanksitem
Y.-F. Liu is with State Key Laboratory of Scientific and Engineering Computing, Institute of Computational Mathematics and Scientific/Engineering Computing, Academy of Mathematics and Systems Science, Chinese Academy of Sciences, China. E-mail: yafliu@lsec.cc.ac.cn.
}% <-this % stops an unwanted space
%\thanks{Part of this work has been presented at the 13th International Symposium on Modeling and Optimization in Mobile, Ad Hoc and Wireless Networks (WiOpt), Mumbai, India, May 25-29, 2015 \cite{WiOpt}.} 
\thanks{This work is partially supported by the General Research Funds (Project No.: CUHK 14202814) established under the University Grant Committee of the Hong Kong Special Administrative Region, China, and the National Natural Science Foundation of China under Grants 11301516 and 11331012.}
%\thanks{Manuscript received April 19, 2005; revised September 17, 2014.}
}

%\IEEEcompsocitemizethanks{
%\IEEEcompsocthanksitem Q. Ma and J. Huang are with the Department of
%Information Engineering, The Chinese University of Hong Kong. E-mail: \{mq012,\ jwhuang\}@ie.cuhk.edu.hk.
%\IEEEcompsocthanksitem L. Gao is with the School of Electronic and Information Engineering, Harbin Institute of Technology Shenzhen Graduate School. E-mail: 	
%gaolin@hitsz.edu.cn.
%\IEEEcompsocthanksitem Y.-F. Liu is with the State Key Laboratory of Scientific and Engineering Computing, Institute of Computational Mathematics and Scientific/Engineering Computing, Academy of Mathematics and Systems Science, Chinese Academy of Sciences. E-mail: yafliu@lsec.cc.ac.cn.
%}

%\markboth{Journal of \LaTeX\ Class Files,~Vol.~13, No.~9, September~2014}%
%{Shell \MakeLowercase{\textit{et al.}}: Bare Demo of IEEEtran.cls for Computer Society Journals}
%% The only time the second header will appear is for the odd numbered pages
%% after the title page when using the twoside option.

\IEEEcompsoctitleabstractindextext{%
\begin{abstract}
Crowdsourced wireless community networks can effectively alleviate the limited coverage issue of Wi-Fi access points (APs), by encouraging individuals (users) to share their private residential Wi-Fi APs with  others.
In this paper, we provide a comprehensive economic analysis for such a crowdsourced network, with the particular focus on the users' behavior analysis and the community network operator's pricing design.
Specifically, we formulate the interactions between the network operator and users as a two-layer Stackelberg model, where the operator determining the pricing scheme in Layer I, and then users determining their Wi-Fi sharing schemes in Layer II.
First, we analyze the user behavior in Layer II via a two-stage membership selection and network access game, for both small-scale networks and large-scale networks.
Then, we design a partial price differentiation scheme for the operator in Layer I, which generalizes both the complete price differentiation scheme and the single pricing scheme (i.e., no  price  differentiation).
We show that the proposed partial pricing scheme can achieve a good tradeoff between the revenue and the implementation complexity.
Numerical results demonstrate that when using the partial pricing scheme with only two prices, we can increase the operator's revenue up to 124.44\% comparing with the single pricing scheme, and can achieve an average of 80\% of the maximum operator revenue under the complete price differentiation scheme.
\end{abstract}


}

\maketitle

\IEEEdisplaynotcompsoctitleabstractindextext

\IEEEpeerreviewmaketitle

\section{Introduction}\label{sec:intro}

\subsection{Background and Motivation}\label{sec:moti}

Global mobile data traffic grows rapidly nowadays, with an unprecedented anticipated annual growth rate of $57\%$ from 2014 to 2019 \cite{demand}.
The capacity of cellular networks, however, increases much slower than the mobile data traffic.
Hence, Wi-Fi networks are playing an increasingly important role in bridging such a gap by carrying a significant amount of mobile data traffic \cite{app-3,app-4,app-5,app-6,app-7}. 
%\footnote{Typical examples of Wi-Fi applications include mobile data offloading \cite{app-3,app-4} and user-provided networking \cite{app-5,app-6,app-7}.}
%\footnote{According to Cisco's report \cite{demand}, around $46\%$ of the global mobile data traffic was offloaded to the fixed network through Wi-Fi or femtocell in 2014.}
The fast development of Wi-Fi technology is due to several reasons such as low deploying costs of Wi-Fi access points (APs) and high Wi-Fi data transmission rates. 
However, the large-scale deployment of a Wi-Fi network is often restricted by the limited coverage of each single Wi-Fi AP (typically tens of meters indoors and hundreds of meters outdoors \cite{WiFiCoverage}), which is much smaller than the coverage of a cellular tower.
Hence, it is very expensive for a single operator to deploy enough Wi-Fi APs to entirely cover a large area such as a city or a nation.~~~~~~~~~~~~~~~~~~ 

The crowdsourced wireless community network turns out as a promising solution to expand the Wi-Fi coverage with a low cost.
The key idea is to encourage individuals (users) to share their owned private Wi-Fi APs with each other, hence aggregating/crowdsourcing the coverage and capacity of private Wi-Fi APs \cite{WiOpt,communities, motivation,ma-wiopt,Competition,Cooperation,TwoPrice,Graph}.
By crowdsourcing millions of Wi-Fi APs already installed by users, this new type of network can reduce or even remove new Wi-Fi installations for the network operator.
Meanwhile, users can also benefit from joining such a community network, as they can use not only their own APs when staying at home, 
%\footnote{We use ``home'' to denote the location of a user's own Wi-Fi AP, which can correspond to residence, office, or even public areas (such as for those Wi-Fi provided by coffee shops).}, 
but also others' APs when traveling to corresponding locations.
Clearly, the success of such a crowdsourced network largely depends on the active participations and contributions of many individual Wi-Fi owners, and hence requires a careful design of economic incentive mechanism.

One prominent commercial example of wireless community networks is FON \cite{FON}, which has more than 20 millions member Wi-Fi APs globally.
In FON, the operator incentivizes Wi-Fi AP owners to share their private APs by using two different incentive schemes, corresponding to two kinds of memberships: \emph{Linus} and  \emph{Bill}.
As a  {Linus}, a user can get free Wi-Fi access within the community network coverage.
As a  {Bill}, a user can earn money from sharing his AP with other users.
Moreover, a user who does not own a Wi-Fi AP can still access the FON network as an \emph{Alien}, by purchasing Wi-Fi passes from the operator. 
%The success of FON motivates our research in this paper.
Despite the commercial success of wireless community networks, however,   there is little work performing the comprehensive economic analysis for such a new network scheme.

\subsection{Model and Problem Formulation}\label{sec:contri}

\begin{figure}
  \centering
   \includegraphics[width=0.42\textwidth]{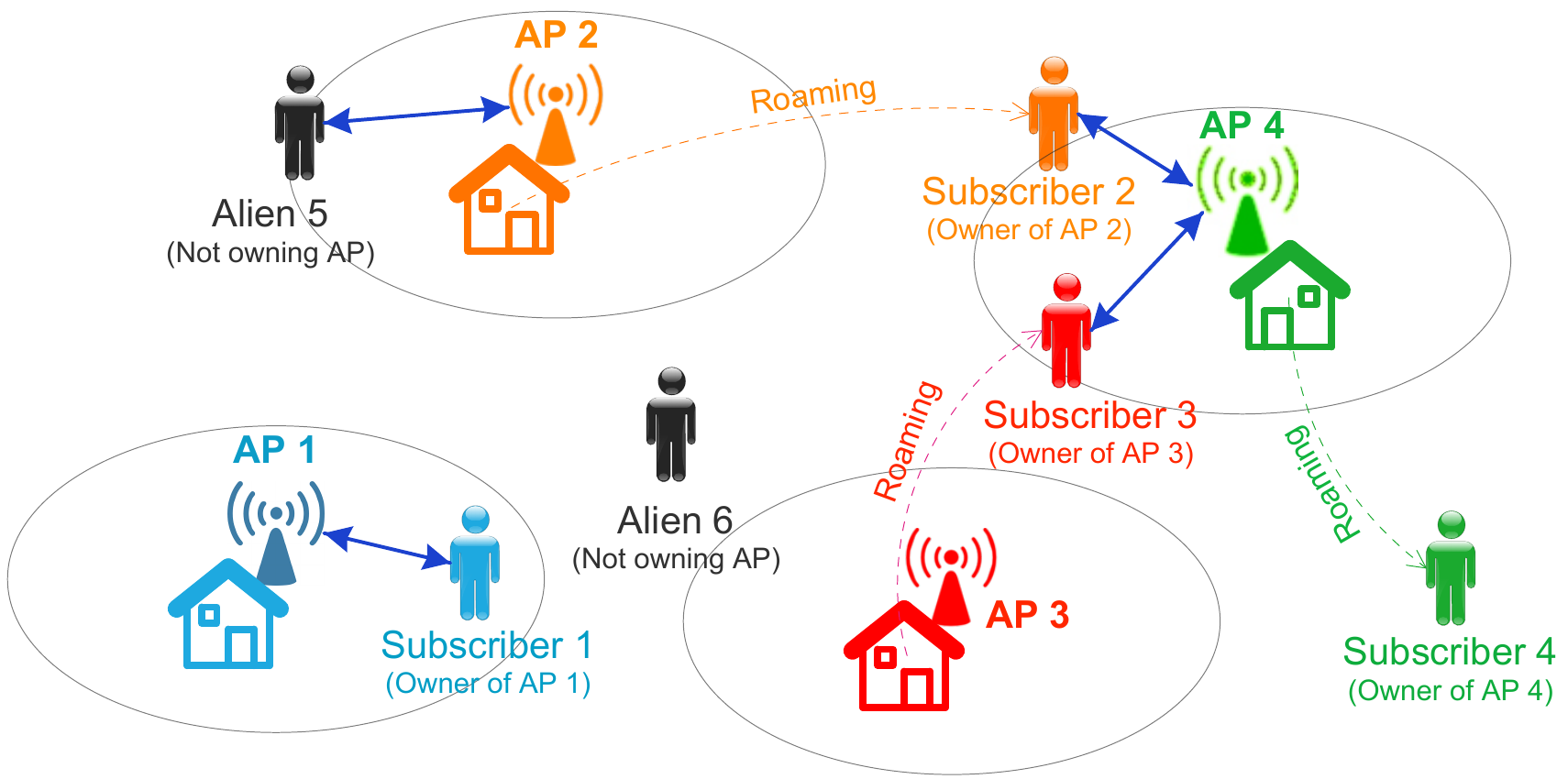}
   \vspace{-3mm}
  \caption{Wireless Community Network Model}\label{fig:model}
  \vspace{-3mm}
\end{figure}

In this work, we consider a wireless community network launched by a FON-like network operator.
As in FON, there are two types of mobile users in the network: \emph{subscribers} and \emph{Aliens}, each traveling (roaming) randomly according to certain mobility
pattern.
Each subscriber owns a private residential Wi-Fi AP at a fixed home location, and opens up his AP for the access of other users.
An Alien does not own a Wi-Fi AP (hence does not contribute to the  network), but can access subscribers' APs (when roaming to the corresponding locations) with a certain fee.
Figure \ref{fig:model} illustrates such a wireless community network, where subscriber 1 (owner of AP 1) stays at home and connects to his own AP, subscribers 2 and 3 travel to subscriber 4's home location and connect to AP 4, and Alien 5  travels to subscriber 2's home location and connects to AP 2.
Subscriber 4 and Alien 6 are at areas without Wi-Fi coverage, hence no available connections.~~~~~~~~

Similar as FON  \cite{FONPass}, the network operator offers two different memberships for subscribers, i.e., \emph{Linus} or \emph{Bill},  corresponding to two different incentive schemes:
\begin{itemize}
  \item As a \emph{Linus}, a subscriber contributes his AP without any monetary compensation, and as compensation, he can use other APs free of charge;
  \item As a \emph{Bill}, a subscriber needs to pay for using other APs (according to a pricing scheme specified by the operator), and can obtain a portion of the revenue collected at his own AP by the network operator.
\end{itemize}
Moreover, an Alien has to pay for using any AP in the network (according to a usage-based pricing scheme specified by the operator), as he does not contribute to the network.
The payments of Alien and Bill (for using other APs) are often \emph{time usage-based}  \cite{FONPass} (i.e., proportional to the Wi-Fi connection time).

The network operator and the users (subscribers and Aliens) interact in the following order.
First, the operator announces the pricing scheme, i.e., the usage-based price (charged to Bills and Aliens) at each AP.
Second, each subscriber chooses his membership type for a given \emph{time period} (e.g., six months), considering his mobility (travel) pattern within that time period as well as his demand and evaluation for network access during travel.
Third, when travelling to a particular AP's location in a particular \emph{time slot} (e.g., five minutes), each user further decides his network access time on that AP in that time slot, taking the network congestion into consideration.
In this work, we will study the above  {operator pricing design} and {user decision problems} systematically.

\begin{figure}
  \centering
   \includegraphics[width=0.32\textwidth]{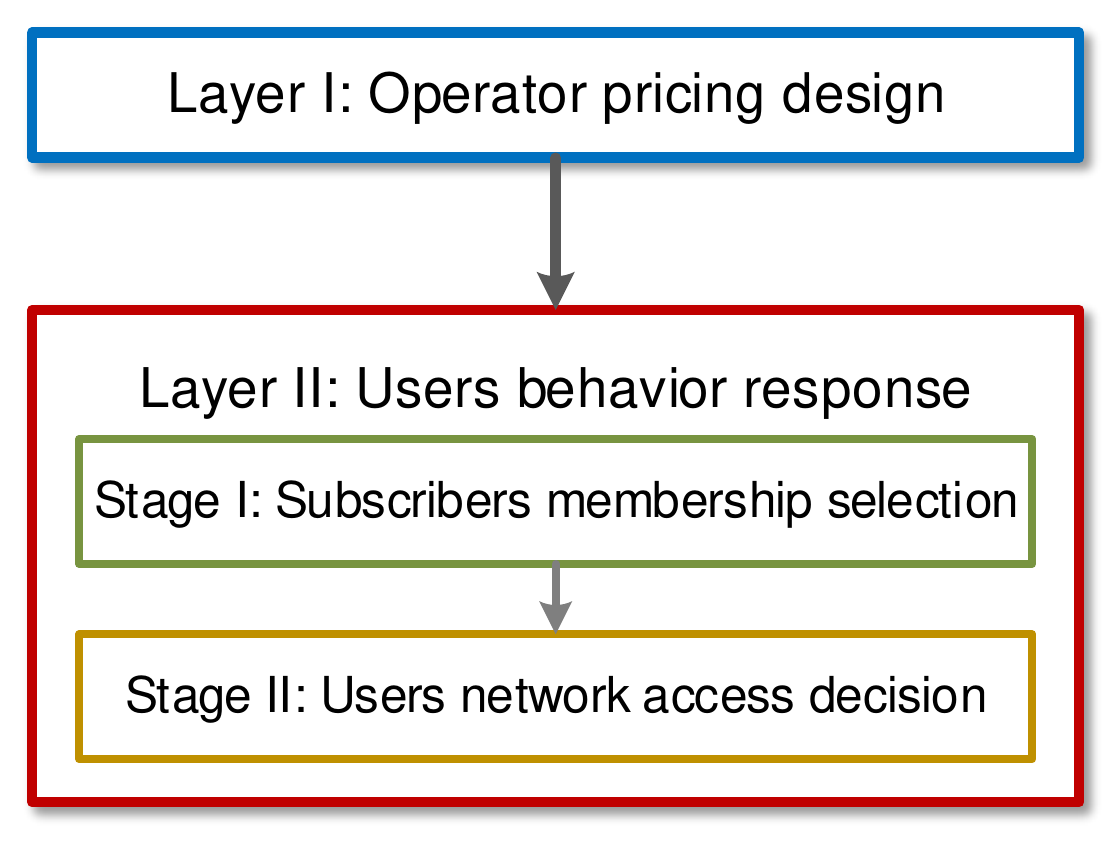}
   \vspace{-3mm}
  \caption{Stackelberg Model of the Operator and Users Interactions}\label{fig:Stackelberg}
\vspace{-3mm}
\end{figure}

\subsection{Solution and Contribution}

We formulate the interactions between the operator and users as a two-layer Stackelberg model, where the operator decides the pricing scheme in Layer I and subscribers decide the membership selection and network access in Layer II,  as illustrated in Figure \ref{fig:Stackelberg}.
We want to emphasize that the operator's pricing decision and the subscribers' membership selection decisions are made once at the beginning of each time period (consisting of many time slots) and remain unchanged during the whole period;
howerver, the subscribers' network access decisions are updated in each time slot, according to their membership selections and the operator's pricing scheme, as well as the real network congestions of their roaming APs.
Next we explain the challenges for solving the two stage problems and our corresponding contributions.

First, we study the operator's decision problem in Layer I.
To extract the maximum surplus from users, the operator can implement a \emph{complete} price differentiation scheme, i.e., charging one price per AP, which will lead to a high implementation complexity and potential customer aversion \cite{PriceDiff}.
To balance the revenue and complexity, we propose a \emph{partial} price differentiation scheme, where APs are divided into multiple groups according to their attributes such as the location popularity and the owner's network access evaluation, and the operator charges different prices to different groups.
Such a partial price differentiation scheme includes both complete price differentiation scheme and single pricing scheme as special cases, but is much challenging to design than the latter two.

Next, we study the user decision problem in Layer II.
As will be shown later in Section \ref{sec:user}, to make the best decision, the computational complexity (for each user) to reach the decision increases exponentially with the number of users.
Thus, in practice, users may not be able to make the fully rational decisions due to the bounded rationality\cite{BoundedRationality},\footnote{Bounded rationality is the concept that the rationality of decision makers is limited by the cognitive limitations of their minds \cite{BoundedRationality}.
%Bounded rationality is the concept that the rationality of decision makers is limited by the cognitive limitations of their minds, i.e., the capacity of human mind to evaluate and process the information that is available, the time available to make decisions, or the information available \cite{BoundedRationality}.
}
especially in a large-scale network with many users (which is more general).
To this end, we propose an \emph{approximate} Stackelberg model for large-scale networks, and understand the system equilibrium when users make their decisions according to approximately best responses.

As far as we know, this is the first work that makes the comprehensive game-theoretic and economic analysis for a crowdsourced wireless community network. % for both small-scale and large-scale networks.
We summarize the key contributions of this work as follows.

\begin{itemize}

\item \emph{Model Novelty:}
Our model is novel and captures several key practical issues, such as user mobility pattern, network access evaluation, demand response,
and network congestion effect.
These issues have not been fully considered before in the context of wireless community networks.

\item \emph{Small-Scale Network Analysis:}
We propose a Stackelberg model to capture the interactions between the operator and users in the crowdsourced wireless community networks for small networks.
We design a partial price differentiation scheme, which includes the complete price differentiation scheme and single pricing scheme as special cases, to help the operator achieve a balance between revenue and implementation complexity.

\item \emph{Large-Scale Network Analysis:}
We propose an approximate Stackelberg model for the large-scale network with a large number of users,  where users are bounded rational due to the limited computation capability and make the approximately best responses. We analyze the user behavior and the operator pricing design in the large-scale network systematically.

\end{itemize}

The rest of the paper is organized as below.
In Section \ref{sec:model}, we present the   model.
In Section \ref{sec:user}, we analyze the users' game in Layer II.
In Section \ref{sec:pricing}, we analyze the operator's pricing design in Layer I.
In Section \ref{sec:approximate}, we analyze the user behavior and operator pricing design in large-scale networks.
We present simulation results and derive engineering insights in Section \ref{sec:simu}, and conclude in Section \ref{sec:conc}.
%Due to space limit, we put some proofs in the online appendix \cite{report}.

\subsection{Literature Review}\label{sec:review}

There are several closely related studies in wireless community networks, regarding incentive issues \cite{motivation}, the network expansion and interactions with traditional ISP \cite{Competition,Cooperation}, and the pricing mechanism design \cite{TwoPrice,Graph}.
Camponovo \emph{et al.} in \cite{motivation} concluded based on surveys that getting free Internet access from other members and revenue sharing are the two main incentives for users to join the FON network in Switzerland.
Manshaei \emph{et al.} in \cite{Competition} modeled a user's payoff as a function of the subscription fee and network coverage, and studied the evolution dynamics of wireless community networks.
Biczok \emph{et al.} in \cite{Cooperation} studied the competition and cooperation among users, wireless community network operator, and ISPs.
Afrasiabi \emph{et al.} in \cite{TwoPrice} proposed a low introductory price policy to promote the service adoption.
Mazloumian \emph{et al.} in \cite{Graph} proposed a fair pricing, which is proportional to users' predicted network coverage.

In this work, we focus on the user behavior analysis and the operator pricing design in a crowdsourced wireless community network.
Neither problem has been systematically studied in the existing literature.
Our model not only captures the Internet access sharing and revenue sharing, but also incorporates the impact of users mobility and the network congestion effect.
This makes our model more comprehensive and practically significant.~~~~~~

\section{System Model}\label{sec:model}

\subsection{The Network Model}

As illustrated in Figure \ref{fig:model}, we consider a crowdsourced wireless community network consisting of a set $\Ks =\{1,\ldots ,K\}$ of \emph{subscriber} (users), each owning a private residential Wi-Fi AP, and a set $\Ka =\{K+1,...,K+K_{\textsc{a}}\}$ of \emph{Alien} (users) without owning any Wi-Fi AP.
Each subscriber is associated with a ``home'', corresponding to the location of his Wi-Fi AP.
Hence, $\mathcal{K}_s$ also represents the set of AP locations.
For convenience, we refer to the set of all subscribers and Aliens as \emph{user set}, denoted by $\Ku = \Ks \bigcup \Ka$.
Subscribers open their private Wi-Fi APs for the access of other users, hence constitute a community network.

Each subscriber can choose two different memberships: Linus and Bill, corresponding to different incentive schemes.
Specifically, as a \emph{Linus}, a subscriber contributes his   AP without any monetary compensation, and can use other APs free of charge. As a \emph{Bill}, a subscriber needs to pay for using other APs, and can obtain a portion of the revenue collected at his own AP by the network operator.
Moreover, each Alien has to pay for using any AP in the network, as he does not contribute to the network.
For clarity, we summarize the properties of these user types in Table \ref{table1}.

We consider the operation in a long time period (e.g., six months), which is divided into $T$ time slots (e.g., five minutes per time slot).
For notational convenience, we normalize the length of each time slot to be one.
We consider a quasi-static mobility model, where each user moves randomly across time slots, and remains at the same location within one time slot.
Let $\eta_{i,j}$, $i\in\Ku, j\in\Ks$ denote the stationary probability that a user $i\in \Ku$ appears at the location of AP $j \in\Ks $ in any time slot, and $\eta_{i,0}$ denote the probability that user $i$ appears at a location that is not covered by any Wi-Fi AP in the community network.
We further define $\boldsymbol{\eta}_i=[\eta_{i,0},\eta_{i,1}, \ldots , \eta_{i,K}]$ as user $i$'s \emph{mobility pattern}. Obviously, $ \sum_{j=0}^K \eta_{i,j} =1 ,\ \forall i\in \Ku$.

\begin{table}[t]
\newcommand{\tabincell}[2]{\begin{tabular}{@{}#1@{}}#2\end{tabular}}
\centering
\caption{A Summary of Three User Types}
\begin{tabular}{|c|c|c|c|}
\hline
\textbf{User Type} & \tabincell{c}{\textbf{Pay for using other APs}} &  \tabincell{c}{\textbf{Paid by sharing his AP}}  \\
\hline
Linus & No & No \\
\hline
Bills & Yes & Yes \\
\hline
Aliens & Yes & Not Applicable  \\
\hline
\end{tabular}
\label{table1}
\vspace{-3mm}
\end{table}

Furthermore, to ensure a subscriber's Quality of Service (QoS) at his home location, we assume that each subscriber splits the bandwidth of his AP into two separate channels (similar as the current practice of FON \cite{FON}):
a \emph{private channel} for serving himself,
and a \emph{public  channel} for serving other users roaming at this location.
Hence, roaming users' communications will not interfere with a subscriber's own communication, and the network congestion only occurs among roaming users on the public channel.

\subsection{Multi-Stage Interactions}

The operator and users interact in the following order.

First, the operator determines and announces the \emph{pricing scheme} at the beginning of the time period, which specifies the Wi-Fi access price on each AP paid by Aliens and Bills (except the AP owner), denoted by $\boldsymbol{p}=\{p_i,\forall i \in \Ks\}$.
The operator's goal is to choose a proper set of prices to maximize the total revenue collected at all member APs.

Second, given the operator's pricing scheme $\boldsymbol{p}=\{p_i,\forall i \in \Ks\}$, each subscriber $i \in \Ks $ chooses his membership $x_i \in \{0, 1\}$ for the entire period of $T$ time slots, where $0$ and $1$ correspond to ``Linus'' and ``Bill'', respectively.
The goal of each subscriber is to choose the best membership that maximizes his expected payoff during the whole period of $T$ time slots, taking his mobility pattern and   demand (or evaluation) for network access as well as other users' membership selections into consideration. % (see Subsection \ref{sec:usa} for more details).

Third, given the operator's pricing scheme $\boldsymbol{p}=\{p_i,\forall i \in \Ks\}$ and the subscribers' membership selections $\boldsymbol{x}=\{x_i, \forall i\in\Ks\}$, each user (subscriber or Alien) further decides the network usage in each time slot, i.e., the \emph{network access time} on the AP of his current location.
When staying at home, a subscriber uses his private channel exclusively, and his network access decision is independent of other users' decisions.
When accessing the Internet through another subscriber's AP, a user (subscriber or Alien) needs to compete for the public channel with other users connecting to the same AP (except the owner of that AP),
hence his best network access decision depends on other users' decisions.

In this work, we will study the operator pricing design problem and the user joint  membership selection and network access decision problem comprehensively.

\subsection{Problem Formulation}

\begin{figure}
  \centering
   \includegraphics[width=0.5\textwidth]{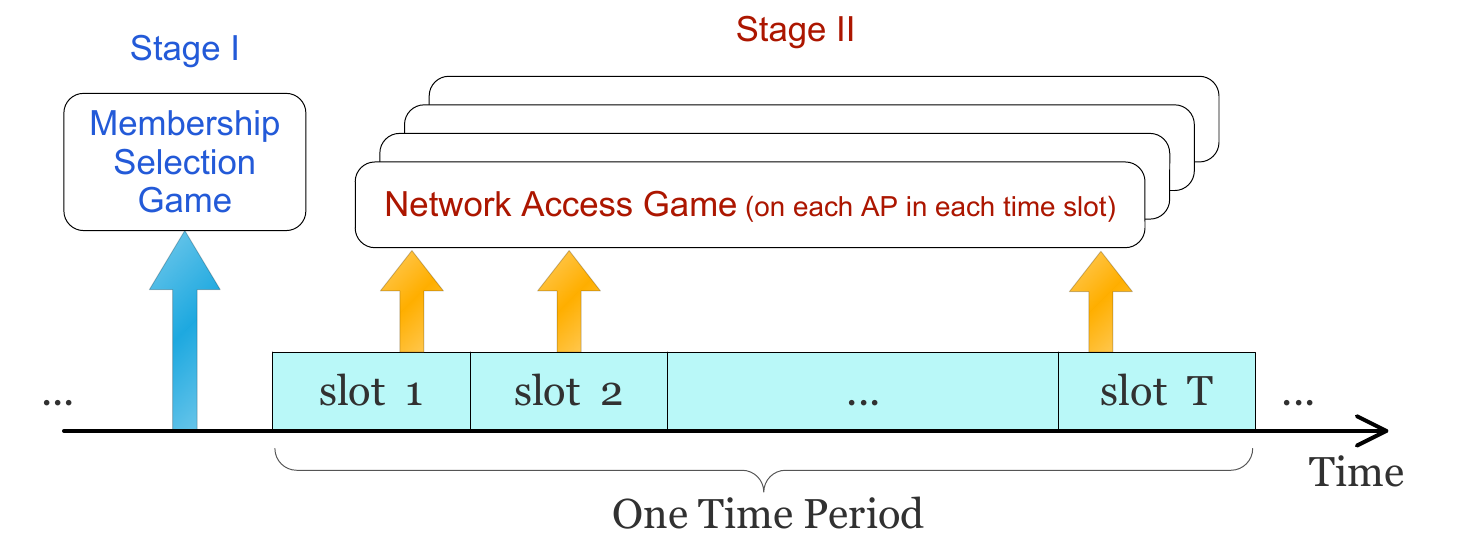}
   \vspace{-5mm}
  \caption{Two-Stage Dynamic Game in Layer II
  (In each time slot, there is a set of parallel network access games, each associated with an AP.)}\label{fig:game}
\vspace{-4mm}
\end{figure}

We formulate the interactions between the operator and the users as a two-layer Stackelberg model, as illustrated in Figure \ref{fig:Stackelberg}.
In Layer I, the operator acts as the \emph{leader} and optimizes the pricing scheme, based on his anticipation of users' responses (i.e., membership selection and network access) to the pricing scheme.
The operator announces the pricing scheme to users at the beginning of each time period, and the pricing scheme will not change throughout  the whole time period.
In Layer II, the users act as the \emph{followers} and decide their membership selections and network access decisions, given the operator's pricing scheme.

Moreover, we formulate users' joint membership selection and network access problem (in Layer II) as a two-stage dynamic game, as illustrated in Figure \ref{fig:game}.
In Stage I, subscribers participate in a \emph{membership selection game} at the beginning of each time period, where each subscriber chooses his membership for the whole time period.
In Stage II, at each time slot, users travelling to the same AP participate in a \emph{network access game}, where each user decides his network access time on that AP.
Namely, each AP is associated with a network access game at each time slot.

In this work, we consider both the small-scale and large-scale networks.
In practice, each user has the limited computation capability, which may be enough for computing his best decision in a small-scale network, while not enough in a large-scale network.
Hence, we assume that users will make the approximately best responses in large-scale networks, and hence are \emph{bounded rational}.
Namely, we propose an approximate Stackelberg model for large-scale networks, where users act in a bounded rational manner.

In what follows, we will study the Stackelberg game by backward induction, starting from the users' two-stage dynamic game in Layer II (Section \ref{sec:user}), and then moving to the operator's pricing design in Layer I (Section \ref{sec:pricing}).
We will further study the approximate Stackelberg model with bounded rational users in Section \ref{sec:approximate}.

\section{Layer II: User Behavior Anslysis} \label{sec:user}

In this section, we study the users' two-stage dynamic game (for small-scale networks) in Layer II, given the operator's pricing scheme.
We analyze the game by backward induction.
We will first analyze the network access game in Stage II for each time slot and then analyze the membership selection game in Stage I for the whole time period.

\subsection{Stage II: Network Access Game on Each AP} \label{sec:usa}

We first study the network access game in Stage II (on each AP in each time slot), given the subscribers' membership selections $\boldsymbol{x}=\{x_i, \forall i\in\Ks\}$ in Stage I and the operator's pricing scheme $\boldsymbol{p}=\{p_i,\forall i \in \Ks\}$ in Layer I.
In this game, each user decides the network access time on the AP at his current location, aiming at maximizing his payoff in the current time slot.

\subsubsection{Network Access Game Formulation}

Without loss of generality, we consider the network access game on a particular AP $k$ in a particular time slot $t$.
Recall that the length of each time slot is normalized to be one for notational convenience.

The \emph{players} of the game are all users traveling to AP $k$ (except the owner of AP $k$) in time slot $t$, denoted by $\K(k, t) = \Ks(k,t) \bigcup \Ka(k,t)$,
where $\Ks(k,t)$ and $\Ka(k,t)$ are the sets of subscribers and Aliens (at AP $k$ and time slot $t$), respectively.
For notational convenience, we will ignore the time index $t$, and hence write the player set as $\K(k) = \Ks(k) \bigcup \Ka(k)$ in the rest of this section,
since we stick on the operations in time slot $t$.

The \emph{strategy} of each player $i \in \K(k)$ is to decide the network access time $\sigma_{i,k} \in [0,1] $ on AP $k$ in time slot $t$.
We denote the strategies of players in $\K(k)$ except $i$ as $\boldsymbol{\sigma}_{-i,k} = \{\sigma_{j,k}, j\neq i, j\in\K(k) \}$.
The \emph{payoff} of player $i $ is a function of both his own strategy $\sigma_{i,k}$ and other players' strategies $\boldsymbol{\sigma}_{-i,k} $, denoted by $v_{i,k}(\sigma_{i,k},\boldsymbol{\sigma}_{-i,k})$ (to be defined later).

More formally, the network access game on a particular AP $k$ (in time slot $t$) and the corresponding Nash equilibrium are defined as follows.

%\vspace{-2pt}
\begin{definition}[Network Access Game on AP $k$]\label{RCgame}
~
\begin{itemize}
\item Players: the set $ \K(k) $ of users traveling to AP $k$;
\item Strategies: the network access time $\sigma_{i,k} \in [0,1]$ of each user $ i \in \K(k) $ on AP $k$;
\item Payoffs: $v_{i,k}(\sigma_{i,k},\boldsymbol{\sigma}_{-i,k})$, $\forall  i \in \K(k) $.
\end{itemize}
\end{definition}

\begin{definition}[Nash Equilibrium]\label{NEII}
A Nash equilibrium of the Network Access Game on AP $k$ (in time slot $t$) is a profile $\boldsymbol{\sigma}_k^\ast=\{\sigma_{i,k}, \forall i\in \K(k)\}$ such that for each user $ i \in \K(k) $,
%\vspace{-2pt}
\begin{equation*}
v_{i,k}(\sigma_{i,k}^\ast,\boldsymbol{\sigma}_{-i,k}^\ast) \geq  v_{i,k}(\sigma_{i,k},\boldsymbol{\sigma}_{-i,k}^\ast),\quad \forall \sigma_{i,k} \in [0,1].
%\vspace{-1pt}
\end{equation*}
\end{definition}

Note that the Nash equilibrium $\boldsymbol{\sigma}_k^\ast$ depends on the player set $\K(k)$, hence can be written as $\boldsymbol{\sigma}_k^\ast (\K(k))$.

\subsubsection{Utility and Payoff Definition}

Before analyzing the Nash equilibrium, we first define users' utility and payoff functions.

\textbf{Utility:}
The \emph{utility} captures a user's satisfaction for accessing the Internet for a certain amount of time.
Due to the diminishing marginal returns principle \cite{logutility, logutility2}, we assume the utility function is increasing and concave.
As a concrete example, we define the utility of user $i\in \K(k)$ on AP $k$ as
\vspace{-2pt}
\begin{equation}
u_i(\sigma_{i,k},\boldsymbol{\sigma}_{-i,k}) = \rho_i \log(1+\bar{r}_{i,k}(\boldsymbol{\sigma}_{-i,k}) \cdot  \sigma_{i,k}), \label{uti}
\vspace{-1pt}
\end{equation}
where $\rho_i$ is user $i$'s \emph{network access evaluation}, characterizing user $i$'s valuation of data consumption.
Here, $\bar{r}_{i,k}(\boldsymbol{\sigma}_{-i,k})$ is the expected data rate that user $i$ can achieve on AP $k$, which is a decreasing function of other users' network access times $\boldsymbol{\sigma}_{-i,k}$ on AP $k$.
Intuitively, with more users accessing AP $k$'s public channel simultaneously, user $i$'s achieved data rate will decrease due to the increased congestion.
Obviously, $\bar{r}_{i,k}(\boldsymbol{\sigma}_{-i,k}) \cdot  \sigma_{i,k} $ denotes the total expected data amount that user $i$ consumes on AP $k$ (in time slot $t$).

\begin{figure} 
  \centering
   \includegraphics[width=0.3\textwidth]{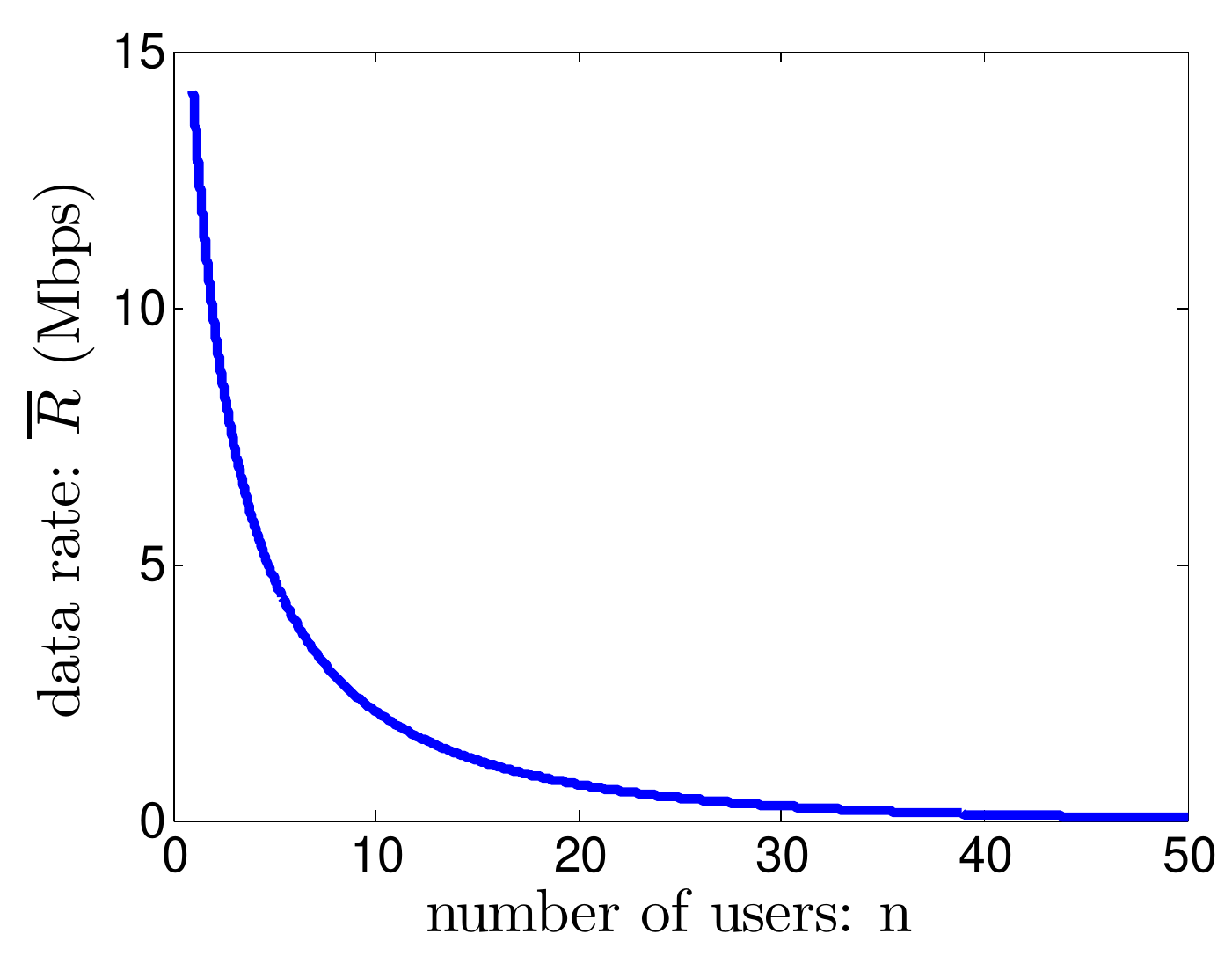}
  \caption{Average Data Rate per Wi-Fi User \cite{20Q}}
  \label{fig:DataRate}
  \vspace{-3mm}
\end{figure}

Next, we derive the concrete form of the user $i$'s expected data rate  $\bar{r}_{i,k}(\boldsymbol{\sigma}_{-i,k})$ on AP $k$.
Let $\bar{R}(n)$ denote the average data rate of a Wi-Fi user when $n$ users are connecting to the Wi-Fi AP simultaneously.
Let $P_{i,k} (n )$ denote the probability that $n $ other users (except $i$) connect to AP $k$.
Then,  user $i$'s expected data rate $\bar{r}_{i,k}(\boldsymbol{\sigma}_{-i,k})$  can be calculated as follows:
\vspace{-2pt}
\begin{equation}
\bar{r}_{i,k}(\boldsymbol{\sigma}_{-i,k})=\sum_{n=0}^{|\K(k)|-1} P_{i,k} (n) \cdot \bar{R}(n+1). \label{DataRate}
\vspace{-2pt}
\end{equation}
According to IEEE $802.11$g standard \cite{20Q}, we have:
\vspace{-2pt}
\begin{equation}
\bar{R}(n)=
\frac{\tau \bar{\tau}^{n-1}L}{\bar{\tau}^nT_b+[(1-\bar{\tau}^n)-n\tau\bar{\tau}^{n-1}]T_c+n\tau\bar{\tau}^{n-1}T_s} , \label{eq:R}
\vspace{-2pt}
\end{equation}
where $\tau$ is the average successful probability of contention (and $\bar{\tau} = 1-\tau$), $L$ is the average payload length, $T_b$ is the length of a backoff slot, $T_c$ is the length of a collision slot, and $T_s$ is the length of a successful slot.
Figure \ref{fig:DataRate} illustrates an example of $\bar{R}(\cdot)$ under IEEE $802.11$g standard (reproduced from \cite{20Q}, with parameters $\tau=0.0765, L=8192, T_b=28\mu s$, and $T_c=T_s=85.7+L/54 \mu s$).
The decreasing data rate per user is due to both the reduced resource per user and the waste of resources caused by congestion among users.

For simplicity, we assume that if a user $i$ decides to access the channel with a certain time $\sigma_{i,k}$, he will spread this access time \emph{randomly and uniformly} across the entire time slot.
Recall that the length of a time slot is normalized to $1$.
Hence, the probability that user $i$ connects to AP $k$
\emph{in an infinitely small time interval within the time slot} is $\sigma_{i,k}$.
Thus, $P_{i,k} (n), n=0,1,...,|\K(k)|-1$, follow the binomial distribution (with a total of $|\K(k)|$ trials and a success probability $\sigma_{j,k}$ for each trial $ j \in \K(k)\setminus \{i\}$).
Formally,
\vspace{-8pt}
\begin{equation*}
%\vspace{-3pt}
P_{i,k} (n ) = \sum_{\K_{n } \in \mathbf{K}_{n }(k)}
\left(
\prod_{j \in \K_{n } } \sigma_{j,k} \cdot  \prod_{j \in \K(k)\setminus\{i\}\setminus\K_{n } } ( 1- \sigma_{j,k})
\right),
%\vspace{-2pt}
\end{equation*}
where $\K_{n } $ denotes an arbitrary subset of $\K(k)$ with $n$ users (except $i$), and $\mathbf{K}_{n }(k)$ denotes the set of all  possible $\K_{n } $.
Obviously, $\prod_{j \in \K_{n } } \sigma_{j,k} $ denotes the probability that all users in $\K_{n }  $ are connecting to AP $k$, and $\prod_{j \in \K(k)\setminus\{i\}\setminus\K_{n } } ( 1- \sigma_{j,k})$ denotes the probability that all other users (except user $i$ and those in $\K_{n } $)  are \emph{not} connecting to AP $k$.

\textbf{Payoff:}
The \emph{payoff} of each user $i\in \K(k)$ is defined as the difference between the utility and the payment.
Specifically, if user $i$ is a Linus (i.e., $i\in \Ks (k)$ and $x_i = 0 $),
he does not need to pay for his network usage on AP $k$.
Hence, the payoff of a Linus-type user $i$ on AP $k$, denoted by $v_{i,k}^\L$, is the same as his utility defined in \eqref{uti}, i.e.,
\begin{equation}
v_{i,k}^\L (\sigma_{i,k},\boldsymbol{\sigma}_{-i,k}) = u_i(\sigma_{i,k},\boldsymbol{\sigma}_{-i,k}). \label{payoffLinus}
\end{equation}
If user $i$ is a Bill (i.e., $i\in \Ks(k)$ and $x_i = 1$) or Alien (i.e., $i\in \Ka (k) $),
he needs to pay for his network usage on AP $k$, and the payment is proportional to his network access time $\sigma_{i,k}$.
Hence, the {payoff} of a Bill-type or Alien user $i$, denoted by $v_{i,k}^\B$, is the difference between utility and payment, i.e.,
\begin{equation}
v_{i,k}^\B(\sigma_{i,k},\boldsymbol{\sigma}_{-i,k}) = u_i(\sigma_{i,k},\boldsymbol{\sigma}_{-i,k})- p_k \sigma_{i,k} . \label{payoffBill}
\end{equation}

Based on the above, we can summarize the payoff of user $i\in \K(k)$ in the Network Access Game (on AP $k$) as follows:
$$
v_{i,k} (\sigma_{i,k},\boldsymbol{\sigma}_{-i,k}) = ~~~~~~~~~~~~~~~~~~~~~~~~~~~~~~~~~~~~~~
$$
\begin{equation}\label{eq:payoff}
\left\{
\begin{aligned}
v_{i,k}^\L (\sigma_{i,k},\boldsymbol{\sigma}_{-i,k}), &\quad \mbox{ if } i \in \Ks(k) \mbox{ and } x_i = 0;
\\
v_{i,k}^\B (\sigma_{i,k},\boldsymbol{\sigma}_{-i,k}), & \quad \mbox{ if } i \in \Ks(k) \mbox{ and } x_i = 1;
\\
v_{i,k}^\B (\sigma_{i,k},\boldsymbol{\sigma}_{-i,k}), & \quad \mbox{ if } i \in \Ka(k).
\end{aligned}
\right.
\end{equation}

\subsubsection{Nash Equilibrium Analysis}

Now we study the Nash equilibrium of the above Network Access Game (on AP $k$).
%Due to space limit, we state most of the detailed proofs in the appendix in the supplemental material section of the manuscript center.

Given all other users' strategies, a user's \emph{best response} is the strategy that maximizes his payoff. The Nash equilibrium is a strategy profile where each user's strategy is the best response to other users' strategies.

\begin{lemma}\label{lemmaBRLinus}
If user $i$ is a Linus, his best response in the Network Access Game on AP $k$ is
\begin{equation}
\sigma_{i,k}^\ast = 1,
\label{BRLinus}
\end{equation}
regardless of other users' strategies.
\end{lemma}

\begin{lemma}\label{lemmaBRBill}
If user $i$ is a Bill or an Alien, his best response in the Network Access Game on AP $k$ is
\begin{equation}
\sigma_{i,k}^\ast = \min\left\{ 1,\max\left\{ \frac{\rho_i}{p_k}-\frac{1}{\bar{r}_{i,k}(\boldsymbol{\sigma}_{-i,k})},0 \right\} \right\},  \label{BRBill}
\end{equation}
which is a function of other users' strategies $\boldsymbol{\sigma}_{-i,k}$.
\end{lemma}

We next illustrate the existence of the Nash equilibrium in the Network Access Game.

\begin{theorem}\label{ExiNEI}
There exists at least one Nash equilibrium in the Network Access Game on AP $k$.
\end{theorem}

Now we discuss the uniqueness of the Nash equilibrium in the Network Access Game on AP $k$.

\begin{proposition}\label{UniNEI}
In a Network Access Game with two players, the Nash equilibrium is unique if
$ \frac{\bar{R}(1)-\bar{R}(2)}{(\bar{R}(2))^2} < 1 .$
\end{proposition}

Note that the condition in Proposition \ref{UniNEI} is always satisfied for practical WiFi systems given in \cite{20Q}.
For the cases with more than two players, however, the uniqueness of   Nash equilibrium depends on   system parameters in a more complicated fashion.
Please refer to the appendix for more detailed discussions.
We further propose a best response update algorithm in the
appendix,
%online appendix \cite{report},
which is guaranteed to linearly converge to the Nash equilibrium under the same condition for the uniqueness of the Nash equilibrium.

\vspace{-1mm}
\subsection{Stage I: Membership Selection Game} \label{sec:membership}

Now we study the subscribers' membership selection game in Stage I, given the operator's pricing scheme $\boldsymbol{p}=\{p_i,\forall i \in \Ks\}$.
In this stage, each subscriber $i\in\Ks$ decides his membership type $x_i \in \{0,1\}$ (i.e., Linus or Bill) at the beginning of the period, aiming at maximizing the overall expected payoff that he can achieve in all $T$ time slots.
Note that an Alien $i \in \Ka$ cannot choose his type, as he has no Wi-Fi AP and does not contribute to the network.

 \vspace{-2mm}

\subsubsection{Membership Selection Game Formulation}

In the Membership Selection Game, \emph{players} are subscribers in the set $\Ks$.
The \emph{strategy} of each player $i\in \Ks$ is to decide his membership $x_i \in \{0, 1\}$, with $x_i = 0$ and $1$ denoting Linus and Bill, respectively.
%Such a membership choice will last for the whole time period.
We denote the strategies of all players except $i$ by $\boldsymbol{x}_{-i} = \{x_j, j\neq i, j \in \Ks\}$.
The \emph{overall payoff} of a player $i$ is sum of the \emph{total expected payoff} on all APs that he may travel to and the \emph{total expected revenue} that he may collect at his own AP (if choosing to be a Bill) during $T$ slots.
It is a function of his own strategy $x_i$ and other players' strategies $\boldsymbol{x}_{-i}$, denoted by $V_i(x_i,\boldsymbol{x}_{-i})$.

Formally, the Membership Selection Game and the corresponding Nash equilibrium are defined as follows. 
Note that the Nash equilibria in Stage II (Definition \ref{NEII}) and Stage I (Definition \ref{NEI}) together form a Subgame Perfect Equilibrium (SPE) of the whole game.

\begin{definition}[Membership Selection Game]\label{MSgame}
$\mbox{ }$
\begin{itemize}
\item Players: the set $\Ks$ of subscribers.
\item Strategies: $x_i\in \{0,1\}$, $\forall i \in \Ks$.
\item Payoffs: $V_i(x_i,\boldsymbol{x}_{-i})$, $\forall i \in \Ks$.
\end{itemize}
\end{definition}

\begin{definition}[Nash Equilibrium]\label{NEI}
A Nash equilibrium of the Membership Selection Game is a profile $\boldsymbol{x}^\ast=\{x_i^\ast,i \in \Ks\}$ such that for each subscriber $i \in \Ks$,
\begin{equation*}
V_i(x_i^\ast,\boldsymbol{x}_{-i}^\ast) \geq  V_i(x_i,\boldsymbol{x}_{-i}^\ast),\quad \forall x_i \in\{0, 1\}.
\end{equation*}
\end{definition}

\subsubsection{Payoff Definition}

Before analyzing Nash equilibrium, we first calculate each subscriber's overall expected payoff in the whole period.

\textbf{Total expected payoff:}
A subscriber's overall expected payoff consists of (i) the total expected payoff on all APs that he may travel to and (ii) the total expected revenue that he may collect on his own AP (if choosing to be a Bill). 
We first calculate the total expected payoff of each subscriber (on all APs that he may travel to), which depends on his mobility pattern.
Recall that the mobility of a subscriber $i$ is characterized by the probabilities of travelling to different APs, i.e., $\boldsymbol{\eta}_i=[\eta_{i,0},\eta_{i,1},\ldots , \eta_{i,K}]$, where $\eta_{i,k} $ is the probability of subscriber $i$ travelling to AP $k$, and $\eta_{i,0}$ is the probability of subscriber $i$ travelling to an area that is not covered by any  AP in the network.
We calculate subscriber $i$'s expected payoffs (per time slot) when staying at home and when roaming outside, respectively.

\emph{(a) When staying at home (with a probability $\eta_{i,i}$)}, subscriber $i$ communicates over the private channel of AP $i$ and does not interfere with other users.
Hence his expected payoff, denoted by $V_{i,i}(x_i,\boldsymbol{x}_{-i})$, is
\vspace{-3pt}
\begin{equation*}
V_{i,i}(x_i,\boldsymbol{x}_{-i}) =  \rho_i \cdot \log(1+\bar{r}_{i,i} \cdot 1),  \label{VkLinus-x}
\vspace{-3pt}
\end{equation*}
where constant $\bar{r}_{i,i} $ corresponds to the average achieved data rate.
The product term $\bar{r}_{i,i} \cdot 1$ implies that user $i$ will access the Internet during the entire time slot.

\emph{(b) When traveling to AP $k \neq i$ (with a probability $\eta_{i,k}$)}, subscriber $i$ needs to compete over the public channel with other users (except $k$) travelling to AP $k$ at the same time (in the Network Access Game).

Suppose that a set $\M(k)$ of other users (except $i$ and $k$) are travelling to AP $k$ at the same time.
That is, the game player set in the Network Access Game on AP $k$ is $\K(k) = \M(k) \bigcup \{i\}$.
For more clarity, let us rewrite the equilibrium payoff of subscriber $i$ on AP $k$, i.e.,  $v_{i,k} (\sigma_{i,k},\boldsymbol{\sigma}_{-i,k})$ defined in \eqref{eq:payoff}, as $v_{i,k} (\sigma_{i,k},\boldsymbol{\sigma}_{-i,k}| \M (k))$, \emph{when competing with a set $\M(k)$ of other users (in the Network Access Game on AP $k$)}.
Hence, the expected payoff of subscriber $i$ on AP $k$ is
\vspace{-3pt}
\begin{equation*}
V_{i,k}(x_i,\boldsymbol{x}_{-i}) =  \sum_{\M(k) \in \mathbf{K}_{-\{i,k\}} } \phi (\M(k)) v_{i,k} (\sigma_{i,k}^\ast,\boldsymbol{\sigma}_{-i,k}^\ast| \M(k)),
\vspace{-3pt}
\end{equation*}
where
$\phi (\M(k))$ is the probability that a set $\M(k)$ of users are travelling to AP $k$,
$(\sigma_{i,k}^\ast, \boldsymbol{\sigma}_{-i,k}^\ast )$ is the corresponding equilibrium of the Network Access Game,
and $\mathbf{K} _{-\{i,k\}} $ is the power set of {$\Ku\setminus \{i,k\}$}, i.e., the set of all subsets of $\Ku\setminus \{i,k\}$.
The probability $\phi (\M(k))$ is given by%\footnote{To calculate $\phi (\M(k))$, a user needs to know the complete information regarding other users' mobility patterns, which may not be realistic in practice. We will study the problem under incomplete information in our future work.}
\vspace{-3pt}
\begin{equation*}
\phi (\M(k)) = \prod_{j \in \M(k)} \eta_{j,k} \cdot \prod_{j \in \Ku \setminus \{i,k\}\setminus \M(k) } (1 - \eta_{j,k}),
\vspace{-3pt}
\end{equation*}
where $\prod_{j \in \M(k)} \eta_{j,k} $ denotes the probability that all users in $\M(k) $ are travelling to AP $k$, and $\prod_{j \in \Ku\setminus \{i,k\}\setminus \M(k)} (1 - \eta_{j,k}) $ denotes the probability that all other users (except users $i$, $k$, and those in $\M(k)$)  are \emph{not} travelling to AP $k$.

\emph{(c) When traveling to an area that is not covered by any AP (with a probability $\eta_{i,0}$)},  the expected payoff of subscriber $i$, denoted by $V_{i,0}(x_i,\boldsymbol{x}_{-i}) $, is%\footnote{If a user can access the Internet through other means, we can normalize the corresponding constant payoff to be zero without affecting the analysis.}
\vspace{-3pt}
\begin{equation*}
V_{i,0}(x_i,\boldsymbol{x}_{-i}) = 0.
\vspace{-3pt}
\end{equation*}

Based on the above, the total expected payoff of subscriber $i$ (on all APs that he may travel to during the whole period of $T$ time slots) is
\vspace{-5pt}
\begin{equation}\label{eq:totalpayoff-1}
V_i^\dag (x_i,\boldsymbol{x}_{-i}) = T \cdot \sum_{k=0}^K
\eta_{i, k} \cdot
V_{i,k}(x_i,\boldsymbol{x}_{-i}) .
\vspace{-3pt}
\end{equation}

\textbf{Total expected revenue:}
Next, we calculate the total expected revenue that each subscriber $i$ may collect on his own AP.
Specifically, if choosing to be a Linus, subscriber $i$ obtains a zero revenue.%\footnote{Note that the operator still charges Bills and Aliens for using a Linus' AP, but does not share the achieved revenue with the Linus.}
If choosing to be a Bill, subscriber $i$ obtains a fixed portion $\delta$ of the revenue collected at his AP.

Suppose that a set $\K(i)$ of other users (except $i$) are travelling to AP $i$. That is, the player set in the Network Access Game on AP $i$ is $\K(i)$.
Then, the Nash equilibrium in the Network Access Game on AP $i$ can be written as
\vspace{-3pt}
\begin{equation}\label{eq:NEAPi}
\{ \sigma_{j,i}^\ast (\K(i)), \forall j\in \K(i)\}.
\vspace{-3pt}
\end{equation}
Recall that the revenue collected on each AP is the total payment of all Aliens and Bills accessing that AP.
Hence, the total revenue collected on AP $i$ is
\vspace{-2pt}
\begin{equation*}%\label{eq:totalrev}
\begin{aligned}
& \Pi_i (\boldsymbol{x}_{-i}, \K(i)) =
\\
 & \sum_{j \in \K(i) \bigcap \Ka} p_i \cdot \sigma_{j,i}^\ast (\K(i))  + \sum_{j \in \K(i) \bigcap \Ks} x_j \cdot p_i \cdot  \sigma_{j,i}^\ast (\K(i)) ,
\end{aligned}
\vspace{-3pt}
\end{equation*}
where the first term is the payment of Aliens, and the second term is the payment of Bills.
Hence, the total expected payment of Bills and Aliens on AP $i$ is
\vspace{-2pt}
\begin{equation*}%\label{eq:exprev}
\bar{\Pi}_i (\boldsymbol{x}_{-i}) = \sum_{\K(i) \in \mathbf{K}_{-i} } \phi (\K(i)) \cdot \Pi_i (\boldsymbol{x}_{-i}, \K(i)) ,
\vspace{-3pt}
\end{equation*}
where $\phi (\K(i)) $ is the probability that a set $\K(i)$ of users are travelling to AP $i$, and $\mathbf{K} _{-i} $ is the power set of $\Ku\setminus \{i\}$.
The probability $\phi (\K(i))$ is given by
\vspace{-2pt}
\begin{equation}\label{eq:phi}
\phi (\K(i)) = \prod_{j \in \K(i)} \eta_{j,i} \cdot \prod_{j \in \Ku \setminus \{i\}\setminus \K(i)} (1 - \eta_{j,i}).
\vspace{-3pt}
\end{equation}

Based on the above, the total expected revenue that a subscriber $i$ can achieve at his own AP (during the whole time period of $T$ time slots) is
\vspace{-2pt}
\begin{equation}\label{eq:totalpayoff-2}
V_i^\ddag (x_i,\boldsymbol{x}_{-i}) = T \cdot x_i \cdot \delta \cdot \bar{\Pi}_i  (\boldsymbol{x}_{-i}) .
\vspace{-2pt}
\end{equation}

\textbf{Overall payoff:}
Combining the total expected payoff in \eqref{eq:totalpayoff-1} and the total expected revenue in \eqref{eq:totalpayoff-2}, the {overall payoff} of each subscriber in the Membership Selection Game is
\vspace{-2pt}
\begin{equation}\label{eq:totalpayoff}
\begin{aligned}
 & V_i (x_i,\boldsymbol{x}_{-i})  = V_i^\ddag (x_i,\boldsymbol{x}_{-i}) +
V_i^\dag (x_i,\boldsymbol{x}_{-i})
\\\vspace{-3pt}
& = T \cdot \left(
x_i \cdot \delta \cdot \bar{\Pi}_i  (\boldsymbol{x}_{-i})
+
\sum_{k=0}^K
\eta_{i, k} \cdot
V_{i,k}(x_i,\boldsymbol{x}_{-i})  \right).
\end{aligned}
\vspace{-3pt}
\end{equation}

\vspace{-3mm}

\subsubsection{Nash Equilibrium Analysis}

A subscriber $i$ will make the membership decision to maximize his overall payoff defined in \eqref{eq:totalpayoff}.
Specifically, he will choose to be a Linus if $V_i(0,\boldsymbol{x}_{-i}) > V_i(1,\boldsymbol{x}_{-i})$, and choose to be a Bill otherwise.
For notational convenience, we denote $f_i(\boldsymbol{x}_{-i})$ as the gap between  $V_i(1,\boldsymbol{x}_{-i}) $ and $ V_i(0,\boldsymbol{x}_{-i})$:
\vspace{-2pt}
\begin{equation}
f_i(\boldsymbol{x}_{-i}) = V_i(1,\boldsymbol{x}_{-i}) - V_i(0,\boldsymbol{x}_{-i}). \label{f}
\vspace{-2pt}
\end{equation}
Hence, subscriber $i$ will choose to be a Linus ($x_i=0$) if $f_i(\boldsymbol{x}_{-i}) < 0$, and choose to be a Bill ($x_i=1$) if $f_i(\boldsymbol{x}_{-i}) \geq  0$.
Mathematically, this is equivalent to choosing $x_i$ from $ \{0 , 1\}$, such that the following condition holds:
\vspace{-2pt}
\begin{equation*}
(2x_i-1)\cdot f_i(\boldsymbol{x}_{-i}) \geq 0.
\vspace{-2pt}
\end{equation*}

Next, we study the Nash equilibrium of the Membership Selection Game.

\begin{lemma}\label{lemmaMS}
A membership profile $\boldsymbol{x}^\ast$ is an Nash equilibrium of the Membership Selection Game, if and only if
\vspace{-2pt}
\begin{equation*}
(2x_i^\ast-1)\cdot f_i(\boldsymbol{x}_{-i}^\ast) \geq 0, \quad  \forall i\in \mathcal{K}.
\vspace{-2pt}
\end{equation*}
\end{lemma}

\begin{proposition}\label{lemmaETA}
For each subscriber $i$, if
\vspace{-2pt}
\begin{equation*}
\eta_{i,i} > \underline{\eta}_{i} \triangleq
1 - \frac{\delta \cdot \bar{\Pi}_i  (\boldsymbol{x}^\ast_{-i})}{
\sum_{k\in \Ks/\{i\}} \left(V_{i,k}(0,\boldsymbol{x}^\ast_{-i}) - V_{i,k}(1,\boldsymbol{x}^\ast_{-i})\right)} ,
\vspace{-2pt}
\end{equation*}
then his best response is to choose to be a Bill (i.e., $x_i=1$).
\end{proposition}

Intuitively, a subscriber with a large probability of staying at home will choose to be a Bill, as his network usage on other APs is small, hence the benefit of obtaining revenue at his own AP outweighs the payment at other APs.

Unfortunately, the Membership Selection Game may not always possess an Nash equilibrium in Definition \ref{NEI}, which is essentially a \emph{pure strategy} equilibrium (i.e., each subscriber chooses a particular membership).
To illustrate this, we provide a simple example with 3 APs in the appendix.
Hence, in what follows, we will further look at the case of \emph{mixed-strategy} Nash equilibrium \cite{Game}, where each subscriber may choose both membership types with probabilities.

\subsubsection{Mixed-Strategy Nash Equilibrium}\label{subsec:mexed}

For each subscriber $i$, his \emph{mixed strategy} can be characterized as the probability $\alpha_i \in [0,1]$ of choosing to be a Bill (hence the probability of choosing to be a Linus is  $1 - \alpha_i$).
The pure strategy $x_i$ is a special case of the mixed strategy when $\alpha_i$ equals 1 or 0.
For writing convenience, we denote the mixed strategy profile of all subscribers except $i$ as
\vspace{-2pt}
\begin{equation*}
\boldsymbol{\alpha}_{-i}=\{\alpha_j,j\neq i, j \in \Ks\}.
\vspace{-2pt}
\end{equation*}
Then, the expected payoff of subscriber $i$ can be defined as
\vspace{-2pt}
\begin{equation}
\omega_i(\alpha_i,\boldsymbol{\alpha}_{-i})=\alpha_i \cdot \bar{V}_i(1,\boldsymbol{\alpha}_{-i})  + (1-\alpha_i) \cdot \bar{V}_i(0,\boldsymbol{\alpha}_{-i}) ,  \label{MixPayoff}
\vspace{-2pt}
\end{equation}
where $\bar{V}_i(1,\boldsymbol{\alpha}_{-i})$ and $\bar{V}_i(0,\boldsymbol{\alpha}_{-i})$ are subscriber $i$'s expected payoffs when choosing to be a Bill and a Linus, respectively.
Note that $\bar{V}_i(1,\boldsymbol{\alpha}_{-i})$ and $\bar{V}_i(0,\boldsymbol{\alpha}_{-i})$ are the expected values over all possible membership selections of all other users.
Specifically, there are $K-1$ other subscribers, hence $2^{K-1}$ possible membership selection combination of those subscribers, forming a set $\mathcal{X}_{-i}$.
Each subscriber $j$ chooses $x_j = 1$ and $0$ with probabilities $\alpha_j$ and  $1 - \alpha_j$, respectively.
Then, the probability that a particular $\boldsymbol{x}_{-i}\in\mathcal{X}_{-i} $ is
\vspace{-2pt}
\begin{equation}
\psi (\boldsymbol{x}_{-i}) = \prod_{j \in \Ks\setminus \{i\}}
\big( \alpha_j \cdot x_j + (1 - \alpha_j) \cdot (1 - x_j) \big). \label{eq:psi}
\vspace{-2pt}
\end{equation}
Then, $\bar{V}_i(1,\boldsymbol{\alpha}_{-i})$ and $\bar{V}_i(0,\boldsymbol{\alpha}_{-i})$ can be calculated by
\vspace{-2pt}
\begin{equation}
\bar{V}_i(x_i,\boldsymbol{\alpha}_{-i})
=
\sum_{\boldsymbol{x}_{-i} \in\mathcal{X}_{-i}}
\psi (\boldsymbol{x}_{-i}) V_i(x_i,\boldsymbol{x}_{-i}) , x_i\in\{0,1\}, \label{eq:ep}
\vspace{-2pt}
\end{equation}
where $V_i(x_i,\boldsymbol{x}_{-i})$ is the overall payoff of subscriber $i$ under the pure strategy profile defined in \eqref{eq:totalpayoff}.

\begin{definition}[Mixed-Strategy Nash Equilibrium]\label{NEImixed}
A mixed-strategy Nash equilibrium of the Membership Selection Game is a probability profile $\boldsymbol{\alpha}^\ast$ such that for each subscriber $i \in \Ks$:
$$ \omega_i(\alpha_i^\ast,\boldsymbol{\alpha}_{-i}^\ast) \geq  \omega_i(\alpha_i,\boldsymbol{\alpha}_{-i}^\ast), \quad \forall \alpha_i\in[0,1]. $$
\end{definition}

We first show the existence of the mixed-strategy Nash equilibrium in the Membership Selection Game.

\begin{theorem}\label{lemmaExiMixedNE}
There exists at least one mixed-strategy Nash equilibrium in the Membership Selection Game.
\end{theorem}

To compute the Nash equilibrium effectively, we design a \emph{smoothed} best response updated algorithm, where each player updates his mixed strategy in a smoothed best response manner according to the other players' mixed strategies in the previous iteration.
The basic idea is as follows.
First, given the mixed strategy profile $\boldsymbol{\alpha}^n$ at the $n$-th round, each player computes the corresponding expected payoff when choosing to be Bill (i.e., $\widetilde{V}_i(1,\boldsymbol{\alpha}_{-i}^n)$) or to be Linus (i.e., $\widetilde{V}_i(0,\boldsymbol{\alpha}_{-i}^n)$).
Then, each player updates his mixed strategy at the $(n+1)$-th round according to the following smoothed best response method \cite{Game-Learning}:
\vspace{-5pt}
\begin{equation*}
\alpha_i^{n+1}=\frac{ e^{  \widetilde{V}_i(1,\boldsymbol{\alpha}_{-i}^n) /\gamma }}{ e^{  \widetilde{V}_i(1,\boldsymbol{\alpha}_{-i}^n) /\gamma }+e^{  \widetilde{V}_i(0,\boldsymbol{\alpha}_{-i}^n) /\gamma}} ,
\vspace{-3pt}
\end{equation*}
where $\gamma $ is a parameter that determines the degree to which the function deviates from the true best response (a larger $\gamma$ implies that the player is more likely to act randomly).
Using the result in \cite{Game-Learning}, we can show that such a smoothed best response with some learning rules (as in fictitious play) converges to the mixed strategy Nash equilibria.

\begin{algorithm}[t]
\caption{Smoothed Best Response Update Algorithm}
\label{algo:br}
\begin{algorithmic}[1]
\REQUIRE
$\boldsymbol{\alpha}^0,\gamma,\varepsilon . $
\ENSURE
$\boldsymbol{\alpha}^\ast . $
\STATE Set $n = 0$ and $Flag = 0$.\\
\WHILE {$Flag = 0$}
\FOR{$i=1:K$}
\STATE Calculate $\widetilde{V}_i(1,\boldsymbol{\alpha}_{-i}^n)$ and $\widetilde{V}_i(0,\boldsymbol{\alpha}_{-i}^n)$.
\STATE Update $\alpha_i^{n+1}=\frac{ e^{  \widetilde{V}_i(1,\boldsymbol{\alpha}_{-i}^n) /\gamma }}{ e^{  \widetilde{V}_i(1,\boldsymbol{\alpha}_{-i}^n) /\gamma }+e^{  \widetilde{V}_i(0,\boldsymbol{\alpha}_{-i}^n) /\gamma}} $.
\ENDFOR
\IF {$|\boldsymbol{\alpha}^{n+1}-\boldsymbol{\alpha}^n|\leq \varepsilon$}
\STATE Set $Flag = 1$.
\ENDIF
\STATE Set $n = n+1$.
\ENDWHILE
\STATE Set $ \boldsymbol{\alpha}^\ast=\boldsymbol{\alpha}^n. $
\end{algorithmic}
\end{algorithm}

\section{Layer I: Operator Pricing Design} \label{sec:pricing}

In this section, we study the operator's pricing design  (for small-scale networks) in Layer I.
The operator designs the pricing scheme based on his anticipation of users' membership selections and network access decisions in Layer II.

Note that the operator can choose different pricing schemes. Two typical examples are the complete price differentiation scheme, where the operator charges different prices on different APs, and the single-pricing scheme, where the operator charges the same price on all APs.
The former scheme can achieve a high performance (operator revenue) with the cost of a high computational complexity, while the latter scheme reduces the complexity with the cost of revenue loss.
To balance the complexity and performance, we will propose a partial price differentiation scheme, which generalizes both the complete price differentiation scheme and the single-pricing scheme.

\subsection{Complete Price Differentiation}

We first study the complete price differentiation scenario, where the operator charges different prices on different APs.
The operator optimizes his pricing scheme to maximize his total expected revenue in one time period, which is proportional to the network usage of Bills and Aliens.
For convenience, we denote the network usage of Bills and Aliens as the \emph{charged} network usage, which brings revenue for the operator directly.

\textbf{Charged network usage}:
We first derive the charged network usage on a particular AP $i \in \Ks$ in a particular time slot, which depends on the set of users that are traveling to AP $i$.
Suppose that a set $\K(i)$ of users (except $i$) are traveling to AP $i$ in that time slot.
That is, the player set in the corresponding network access game is $\K(i)$.
The Nash equilibrium of the game is given in \eqref{eq:NEAPi},
from which we can further compute the charged network usage on AP $i$.
Hence, given the user set $\K(i)$, the charged network usage on AP $i$ in one time slot is:
$$
\sum_{j \in \K(i)\cap\Ka} \sigma_{j,i}^\ast(\K(i))+\sum_{j\in\K(i)\cap\Ks}x_j \cdot \sigma_{j,i}^\ast(\K(i)),
$$
where $\sigma_{j,i}^\ast$ is user $j$'s equilibrium network usage on AP $i$, which is given in Lemmas \ref{lemmaBRLinus} and \ref{lemmaBRBill}.
The first term denotes the total network usage from Aliens in $\K(i)$, and the second term denotes the total network usage from Bills in $\K(i)$.

\begin{figure*}
%\vspace{-5mm}
\begin{align}
& \bar{\sigma}_i^\ast (\boldsymbol{p})=
T \cdot \sum_{\boldsymbol{x}_{-i}\in \mathcal{X}_{-i}} \psi(\boldsymbol{x}_{-i}) \left[ \sum_{\K(i)\in \mathbf{K} _{-i}}  \phi(\K(i)) \left( \sum_{j \in \K(i)\cap\Ka} \sigma_{j,i}^\ast(\K(i))+\sum_{j\in\K(i)\cap\Ks}x_j \sigma_{j,i}^\ast(\K(i)) \right)   \right].\label{eq:EXPusa}
\end{align}
\hrulefill \vspace*{4pt}
%\vspace{-3mm}
\end{figure*}

Next we derive the charged network usage on AP $i$ in the whole period of $T$ time slots,
which is the expected value over all possible membership selection combinations of the other $K-1$ subscribers (except $i$).\footnote{User $i$'s own membership selection does not affect the charged network usage on AP $i$, as each AP owner uses his private channel exclusively and does not account for the charged network usage.}
For notational convenience, we use $\boldsymbol{x}_{-i}=\{x_j:j\neq i, j\in \Ks\}$ to represent a membership selection combination of other $K-1$ subscribers, and use $\mathcal{X}_{-i}$ to represent the set of all $2^{K-1}$ possible membership selection combinations.
Then, the expected charged network usage on AP $i$ in the whole time period, denoted as $\bar{\sigma}_i^\ast (\boldsymbol{p})$, can be calculated by Eq. \eqref{eq:EXPusa} on the top of next page,
where $\mathbf{K}_{-i} $ is the power set of $\Ku\setminus \{i\}$.
The term $\phi(\K(i))$ denotes the probability that a set $\K(i)$ of users are travelling to AP $i$ in a time slot, and is calculated by Eq. \eqref{eq:phi}.
The term $\psi(\boldsymbol{x}_{-i})$ denotes the probability that a particular membership selection combination $\boldsymbol{x}_{-i}\in \mathcal{X}_{-i}$ occurs, and is calculated by Eq. \eqref{eq:psi}.

\textbf{Operator's expected revenue}:
The total expected revenue generated on AP $i$ is proportional to the charged network usage, i.e., $p_i \cdot \bar{\sigma}_i^\ast (\boldsymbol{p})$, where $p_i$ is the price per unit connection time charged to Bills and Aliens on AP $i$.
We use $\alpha_i^\ast$ to represent subscriber $i$'s probability of choosing to be a Bill at equilibrium.
Recall that if subscriber $i$ chooses to be a Linus (with probability $1-\alpha_i^\ast$), the operator can obtain all the revenue generated on AP $i$.
If subscriber $i$ chooses to be a Bill (with probability $\alpha_i^\ast$), the operator can only obtain $1-\delta$ of the revenue generated on AP $i$.
Hence, the operator's expected revenue collected on AP $i$ is:
\begin{equation}
h_i(\boldsymbol{p})=\big[ 1 \cdot \left( 1-\alpha_i^\ast \right) + \left( 1-\delta \right) \cdot \alpha_i^\ast \big] \cdot  p_i \cdot \bar{\sigma}_i^\ast (\boldsymbol{p}).
\end{equation}
Thus, the operator's total expected revenue collected on all APs in the whole time period is
\begin{equation*}
H^C(\boldsymbol{p})=\sum_{i=1}^K h_i(\boldsymbol{p}).
\end{equation*}

Based on the above analysis, the operator's complete price differentiation problem can be defined as follows:
\begin{align}
& \mbox{\textbf{Problem 1: Complete Price Differentiation}} \notag\\
& \displaystyle \max ~~ \displaystyle
H^C(\boldsymbol{p}) = \sum_{i=1}^{K} h_i(\boldsymbol{p}) \notag\\
& \mbox{ var:} ~~~  p_i  \geq 0, \forall i \in \Ks \notag
\end{align}

We can easily compute an effective upper bound for each price variable $p_i$ in the above Problem 1:
\begin{equation}\label{pu}
p_i \leq \bar{p} \triangleq \max_{i \in \Ku}\ \rho_i \cdot \bar{R}(1),
\end{equation}
where $\rho_i$ is the network access evaluation of user $i$ (characterizing user $i$'s valuation of data consumption),
and $\bar{R}(1)$ is the average data rate that a user can achieve when accessing a Wi-Fi AP exclusively, and can be calculated by Eq. \eqref{eq:R}.
By Lemma \ref{lemmaBRBill}, we can easily find that any price $p_i$ larger than $\bar{p}$ will lead to a zero network usage for all Bills and Aliens on AP $i$, hence a zero revenue for the operator.
Thus, we can focus on finding the optimal prices within their respective upper bounds, without affecting the optimality.
Moreover, such upper bounds are also very useful for implementing the iterative algorithm to solve Problem 1 numerically, which will be discussed soon later.

Unfortunately, it is very challenging to solve Problem 1, as it is a mixed multi-level problem, where continuous and binary variables are coupled with each other in a highly nonlinear manner.
Nevertheless, we can explore some useful characteristics of Problem 1, which will help us find the solution numerically.

First, the derivative of the objective function is not readily computable, since we do not have the explicit function relationship of the membership profiles with respect to the prices.
Second, given any price vector $\boldsymbol{p}$, we can compute the objective value, since the optimal total charged network usage on each AP $\bar{\sigma}_i^\ast (\boldsymbol{p})$ is given by Eq. \eqref{eq:EXPusa}, and the equilibrium $\boldsymbol{\alpha}^\ast(\boldsymbol{p})$ of Membership Selection Game can be computed by the smoothed best response update algorithm.
The unavailability of the derivative information of the objective function makes the use of gradient-based methods impossible, and availability of the objective value makes the use of derivative-free algorithms possible.
These factors motivates us to use the derivative-free algorithm \cite{DeriFree}.

We propose to use a recently developed DYCORS (DYnamically COordinate search using Response Surface models) algorithm \cite{DYCORS}, which is one of the derivative-free algorithms, to solve Problem 1.
A DYCORS algorithm is often designed to solve the box-constrained optimization problem.
The key idea of   DYCORS   is to build and maintain a surrogate model \cite{Surrogate} of the objective function at each iteration, and generate trial solutions by using a dynamic coordinate search strategy.
It selects the iterate from a set of random trail solutions obtained by perturbing only a subset of the coordinates of the current best solution, which is helpful in finding the global minimum.
Moreover, the probability of perturbing a coordinate decreases as the algorithm reaches the computational budget.
If the objective function of Problem 1 is continuous, then the DYCORS algorithm converges to a global optimal solution with probability one. 
%For details about DYCORS algorithm, please refer to \cite{DYCORS}.

\subsection{Partial Price Differentiation}

The complete price differentiation is of high implementation complexity and user aversion.
However, if the operator implements the single pricing scheme and ignores the difference among different APs, it may suffer from a high revenue loss.
To this end, we propose a partial price differentiation scheme, where the operator charges the same price on APs with similar attributes, to achieve a tradeoff between the revenue and implementation complexity.
Note that the complete price differentiation scheme and the single pricing scheme are special cases of the partial price differentiation scheme.

In the partial price differentiation scheme, the operator charges the same price on the APs in the same group (i.e., those with similar attributes), while different prices on different groups.
This is analogous to the third-degree price discrimination in economics \cite{ThirdDegreePD, lin}, where prices are set according to user segmentation (based on user attributes such as ages, occupations, and genders).

In our partial price differentiation, all APs are first segmented into different groups, based on the AP attributes (such as location hotness, i.e., the summation of other users' probabilities of roaming to that location, also called location popularity) and the AP owners' network access evaluation.
Then, a single price is set for all APs in the same group.

\subsubsection{AP Segmentation}\label{sec:APsegmentation}

We assume that the operator segments the set $\Ks=\{1,2,\cdots, K\}$ of APs into $G ~(G\leq K)$ groups.
We denote the set of APs in group $g ~(g=1,2,\cdots,G)$ as set $\mathcal{S}_g$,
and we denote the AP segmentation result, i.e., the set of all $G$ AP sets, as $\boldsymbol{\mathcal{S}}=\{\mathcal{S}_1,\mathcal{S}_2,\cdots,\mathcal{S}_G\}$.
The AP segmentation is based on AP attributes.
The APs in the same set $\mathcal{S}_g$ have similar attributes.

The AP segmentation problem has several characteristics.
First, the number of APs involved in the AP segmentation problem can be very large, e.g., several hundred or several thousand.
Second, the AP segmentation is based on AP attributes, such as location hotness and AP owner's network access evaluation.
So each AP is described by a multiple dimensional data profile in the AP segmentation problem.
Third, different AP attributes can be of different importance in the segmentation problem.
The above three characteristics motivate us to use the weighted k-means clustering \cite{segmentation, weightkMeans} to deal with the AP segmentation problem.

The k-means clustering is popular for many real world applications such as user segmentation in marketing research \cite{segmentation}.
The weighted k-means clustering \cite{weightkMeans} assigns a weight to each attribute, where the weight measures the importance of the weight in the clustering.
Details of the weighted k-means clustering can be found in the appendix.

By applying the weighted k-means clustering, we can obtain the AP segmentation result $\boldsymbol{\mathcal{S}}=\{\mathcal{S}_1,\mathcal{S}_2,\cdots,\mathcal{S}_G\}$.

In Section \ref{sec:simu}, we provide numerical results to illustrate how the AP attributes and corresponding weights affect the AP segmentation result.

\subsubsection{Price Optimization}

In the partial price differentiation scheme, the operator sets one price to APs in the same group.

Given the AP segmentation $\boldsymbol{\mathcal{S}}=\{\mathcal{S}_1,\mathcal{S}_2,\cdots,\mathcal{S}_G\}$, the operator's partial price differentiation problem is as follows:
\begin{align}
& \mbox{\textbf{Problem 2:  Partial Price Differentiation}} \notag\\
& \displaystyle \max ~~ \displaystyle
H^P(\boldsymbol{p}^P) = \sum_{i=1}^{K} h_i(\boldsymbol{p}^P) \notag\\
& \mbox{ var:} ~~~ p_g^P  \geq 0, \forall g=1,2,\cdots, G \notag
\end{align}
Here, $\boldsymbol{p}^P=\{p_1^P,p_2^P,\cdots,p_G^P\}$.
The objective function $H^P(\boldsymbol{p}^P)$ is similar as $H^C(\boldsymbol{p})$, by replacing $p_i$ in $H^C(\boldsymbol{p})$ with $p_g^P$ if $i\in \mathcal{S}_g$.
Obviously, Problem 2 has the similar structure as Problem 1, hence we can use the DYCORS algorithm to solve it.

\section{Approximate Model and Analysis for Large-Scale Systems} \label{sec:approximate}

It is important to note that the complexity of the Stackelberg model (in Sections \ref{sec:user} and \ref{sec:pricing}) increases exponentially with the number of APs.
Hence, in large-scale networks with a large number of APs, users may not be able to make the fully rational decisions due to the limited computation abilities.
In this section, we will propose and analyze an approximate Stackelberg model for such a large-scale system.

The basic framework of the approximate Stackelberg model is similar as that of the Stackelberg model in Figure \ref{fig:Stackelberg}.
Namely, the operator optimizes the pricing scheme in Layer I, and users decide their probabilities of choosing to be Bills/Linues and their network access in Layer II.
The key difference is summarized as follows.
In a large-scale network, each user may \emph{not} be able to enumerate all possible behaviors of other users and then calculate his payoff by Eqs. \eqref{MixPayoff}-\eqref{eq:ep} accordingly, due to the limited computation capability of each user.
Hence, in the approximate Stackelberg model, each user will behave based on the  estimated \emph{expected} behaviors of other users.
In the previous Stackelber model, however, each user will behave based on the complete enumeration of all possible behaviors of all other users.

Next, we will analyze the approximate Stackelberg model by backward induction.

\subsection{Layer II: User Behavior Analysis}

Users' approximate two-stage dynamic game is the same as the one illustrated in Figure \ref{fig:game}.
Subscribers participate in an approximate membership selection game in Stage I and decide their probabilities of choosing to be Bills/Linus at the beginning of one time period.
Users participate in an approximate network access game in Stage II and decide their network access time on the APs in each time slot.
We analyze the approximate two-stage dynamic game by backward induction.

\subsubsection{Stage II: Approximate Network Access Game on Each AP}

The approximate network access game in the approximate Stackelberg model is different from the network access game in the Stackelberg model.
In the large-scale system, when a user travels to an AP location, he can neither derive the set of other users who travel to the same AP location, nor those users' precise network access decisions.
Hence, we assume that each user only knows other users' \emph{expected} network access decisions.

We define the approximate network access game on AP $k$ in the large-scale system as follows.
\begin{game}[Approximate Network Access Game on AP $k$]
~
\begin{itemize}
\item Players: the set $ \Ku=\Ks \cup \Ka $ of users in the network;
\item Strategies: $\tilde{\sigma}_{i,k} \in [0,1]$, $\forall  i \in \Ku $;
\item Payoffs: $\tilde{v}_{i,k}(\tilde{\sigma}_{i,k},\tilde{\boldsymbol{\sigma}}_{-i,k})$, $\forall  i \in \Ku $.
\end{itemize}
\end{game}

The payoff is the difference between the utility and the payment (charged to Bills and Aliens).

The utility of user $i \in \Ku$ on AP $k$ is¡ê¡Ò
\begin{equation}
\tilde{u}_i(\tilde{\sigma}_{i,k},\tilde{\boldsymbol{\sigma}}_{-i,k}) = \rho_i \log(1+\tilde{r}_k(\tilde{\boldsymbol{\sigma}}_{k}) \cdot  \tilde{\sigma}_{i,k}). \label{NewUti}
\end{equation}
Here $\tilde{r}_k(\tilde{\boldsymbol{\sigma}}_{k})$ is the data rate that user $i$ can achieve on AP $k$.
When the number of user in the network is large, we assume that each user will achieve the same data rate $\tilde{r}_k(\tilde{\boldsymbol{\sigma}}_{k})$ on AP $k$, which is a function of all users' network access decisions $\tilde{\boldsymbol{\sigma}}_{k}=\{\tilde{\sigma}_{i,k},\forall i \in \Ku\}$:
\begin{equation}
\tilde{r}_k(\tilde{\boldsymbol{\sigma}}_{k})=\bar{R}(\sum_{i\neq k}\eta_{i,k}\tilde{\sigma}_{i,k}). \label{NewDataRate}
\end{equation}
Here $\sum_{i\neq k}\eta_{i,k}\tilde{\sigma}_{i,k}$ represents the expected number of users who access AP $k$ simultaneously in a single time slot, and $\bar{R}(\sum_{i\neq k}\eta_{i,k}\tilde{\sigma}_{i,k})$ is given by Eq. \eqref{eq:R} and denotes the average data rate of a user when $\sum_{i\neq k}\eta_{i,k}\tilde{\sigma}_{i,k}$ users are connecting to AP $k$ simultaneously.

If user $i$ is a Linus, his payoff is just his utility, since he does not need to pay for his network usage on AP $k$:
\begin{equation}
\tilde{v}_{i,k}^\L (\tilde{\sigma}_{i,k},\tilde{\boldsymbol{\sigma}}_{-i,k}) = \tilde{u}_i(\tilde{\sigma}_{i,k},\tilde{\boldsymbol{\sigma}}_{-i,k}). \label{newPayoffLinus}
\end{equation} 
If user $i$ is a Bill or Alien, his payoff is the difference between his utility and payment:
\begin{equation}
\tilde{v}_{i,k}^\B(\tilde{\sigma}_{i,k},\tilde{\boldsymbol{\sigma}}_{-i,k}) = \tilde{u}_i(\tilde{\sigma}_{i,k},\tilde{\boldsymbol{\sigma}}_{-i,k})- \tilde{p}_k \tilde{\sigma}_{i,k} . \label{NewPayoffBill}
\end{equation}
Here $\tilde{p}_k$ is the unit price charged to Bills and Aliens on AP $k$ in the approximate Stackelberg model.

We summarize the payoff of user $i\in \Ku$ in the Approximate Network Access Game (on AP $k$) as follows:
$$
\tilde{v}_{i,k} (\tilde{\sigma}_{i,k},\tilde{\boldsymbol{\sigma}}_{-i,k}) = ~~~~~~~~~~~~~~~~~~~~~~~~~~~~~~~~~~~~~~
$$
\begin{equation}\label{eq:NewPayoff}
\left\{
\begin{aligned}
\tilde{v}_{i,k}^\L (\tilde{\sigma}_{i,k},\tilde{\boldsymbol{\sigma}}_{-i,k}), &\quad \mbox{ if } i \in \Ks \mbox{ and } x_i = 0;
\\
\tilde{v}_{i,k}^\B(\tilde{\sigma}_{i,k},\tilde{\boldsymbol{\sigma}}_{-i,k}), & \quad \mbox{ if } i \in \Ks \mbox{ and } x_i = 1;
\\
\tilde{v}_{i,k}^\B(\tilde{\sigma}_{i,k},\tilde{\boldsymbol{\sigma}}_{-i,k}), & \quad \mbox{ if } i \in \Ka.
\end{aligned}
\right.
\end{equation}

We derive the best response of user $i \in \Ku$ as follows.

\begin{lemma}\label{NewlemmaBRLinus}
If user $i$ is a Linus, his best response in the Approximate Network Access Game on AP $k$ in the large-scale system~is
\begin{equation}
\tilde{\sigma}_{i,k}^{L\ast} = 1,
\label{NewBRLinus}
\end{equation}
regardless of other users' strategies.
\end{lemma}

\begin{lemma}\label{NewlemmaBRBill}
If user $i$ is a Bill or an Alien, his best response in the Approximate Network Access Game on AP $k$ in the large-scale system is
\begin{equation}
\tilde{\sigma}_{i,k}^{B\ast} = \min\left\{ 1,\max\left\{ \frac{\rho_i}{\tilde{p}_k}-\frac{1}{\tilde{r}_{k}(\tilde{\boldsymbol{\sigma}}_{k})},0 \right\} \right\},  \label{NewBRBill}
\end{equation}
which is a function of users' strategy profile $\tilde{\boldsymbol{\sigma}}_{k}$.
\end{lemma}

Regarding the existence of the Nash equilibrium, we have the same conclusion as that in the network access game of the Stackelberg model, i.e., Theorem \ref{ExiNEI} in Section \ref{sec:usa}.
We can use the same best response update algorithm in Section \ref{sec:usa} to derive the Nash equilibrium for the approximate network access game.

\subsubsection{Stage I: Approximate Membership Selection Game}

Now we study the approximate membership selection game in the large-scale system.

The players are the set $\Ks$ of subscribers.
Each subscriber decides his probability of choosing to be a Bill.
Because of the limited computation capability, users are bounded rational.
Each user aims to maximize his \emph{perceived} payoff which is based on the \emph{expected} behaviors of other users.

We define the approximate membership selection game in the large-scale system as follows.
\begin{game}[Approximate Membership Selection Game]\label{NewMSgame}
$\mbox{ }$
\begin{itemize}
\item Players: the set $\Ks$ of subscribers.
\item Strategies: $\tilde{\alpha}_i\in [0,1]$, $\forall i \in \Ks$.
\item Payoffs: $\tilde{\omega}_i(\tilde{\alpha}_i,\tilde{\boldsymbol{\alpha}}_{-i})$, $\forall i \in \Ks$.
\end{itemize}
\end{game}

The perceived expected payoff of subscriber $i$ choosing to be a Bill with probability $\tilde{\alpha}_i$ is:
\begin{equation}
\tilde{\omega}_i(\tilde{\alpha}_i,\tilde{\boldsymbol{\alpha}}_{-i}) = (1-\tilde{\alpha}_i)\widetilde{V}_i^L(0,\tilde{\boldsymbol{\alpha}}_{-i})+\tilde{\alpha}_i \widetilde{V}_i^B(1,\tilde{\boldsymbol{\alpha}}_{-i}),  \label{eq:PercPayoff}
\end{equation}
where $\widetilde{V}_i^L(0,\tilde{\boldsymbol{\alpha}}_{-i})$ is the perceived payoff when subscriber $i$ chooses to be a Linus, and $\widetilde{V}_i^B(1,\tilde{\boldsymbol{\alpha}}_{-i})$ is the perceived payoff when subscriber $i$ chooses to be a Bill.
Due to the limited computation capability of subscriber $i$, subscriber $i$ can not calculate $\widetilde{V}_i^L(0,\tilde{\boldsymbol{\alpha}}_{-i})$ and $\widetilde{V}_i^B(1,\tilde{\boldsymbol{\alpha}}_{-i})$ based on the enumeration of all possible membership selections of other users, as in Eq. \eqref{eq:ep}.
The calculation of $\widetilde{V}_i^L(0,\tilde{\boldsymbol{\alpha}}_{-i})$ and $\widetilde{V}_i^B(1,\tilde{\boldsymbol{\alpha}}_{-i})$ is based on the \emph{expected} behaviors of other users perceived by subscriber $i$.

Specifically, if subscriber $i$ chooses to be a Linus, his perceived payoff $\widetilde{V}_i^L(0,\tilde{\boldsymbol{\alpha}}_{-i})$ includes his perceived utility on all possible APs that he may travel to during the whole time period.

When Linus $i$ stays at home and access his own AP, his perceived payoff is:
$$
\widetilde{V}_{i,i}^L(0,\tilde{\boldsymbol{\alpha}}_{-i})=\rho_i \log(1+\bar{r}_{i,i}\cdot 1).
$$
Recall that $\bar{r}_{i,i}$ corresponds to the average data rate that Linus $i$ achieves at his private channel.
When Linus $i$ travels to AP $k$, his perceived payoff of connecting to AP $k$ is:
$$
\widetilde{V}_{i,k}^L(0,\tilde{\boldsymbol{\alpha}}_{-i})=\rho_i \log(1+\tilde{r}_k(\tilde{\boldsymbol{\sigma}}_{k}) \cdot 1).
$$
When Linus $i$ travels to an area that is not covered by any of the $K$ Wi-Fi APs, his perceived payoff is:
\begin{equation*}
\widetilde{V}_{i,0}^L(0,\tilde{\boldsymbol{\alpha}}_{-i}) = 0.
\end{equation*}

Based on the above, the perceived payoff of Linus $i$ is
\begin{equation}\label{eq:NewtotalpayoffLinus}
\widetilde{V}_i^L(0,\tilde{\boldsymbol{\alpha}}_{-i}) = T \sum_{k=0}^K \eta_{i, k} \widetilde{V}_{i,k}^L(0,\tilde{\boldsymbol{\alpha}}_{-i}).
\end{equation}

If subscriber $i$ chooses to be a Bill, his perceived payoff $\widetilde{V}_i^B(1,\tilde{\boldsymbol{\alpha}}_{-i})$ includes his payoff on all possible APs that he may travel to during the whole time period and the expected revenue that subscriber $i$ collects on his own AP.

Similarly, when Bill $i$ stays at home and access his own AP, his perceived payoff is:
$$
\widetilde{V}_{i,i}^B(1,\tilde{\boldsymbol{\alpha}}_{-i})=\rho_i \log(1+\bar{r}_{i,i}\cdot 1).
$$
When Bill $i$ travels to AP $k$, his perceived payoff of connecting to AP $k$ is:
$$
\widetilde{V}_{i,k}^B(1,\tilde{\boldsymbol{\alpha}}_{-i})=\rho_i \log(1+\tilde{r}_k(\tilde{\boldsymbol{\sigma}}_{k}) \tilde{\sigma}_{i,k}^{B\ast})-\tilde{p}_k\tilde{\sigma}_{i,k}^{B\ast},
$$
where $\tilde{\sigma}_{i,k}^{B\ast}$ is calculated by Eq. \eqref{NewBRBill}.
When Bill $i$ travels to an area that is not covered by any of the $K$ Wi-Fi APs, his perceived payoff is:
\begin{equation*}
\widetilde{V}_{i,0}^B(1,\tilde{\boldsymbol{\alpha}}_{-i}) = 0.
\end{equation*}

The expected revenue that Bill $i$ collects on his own AP is a $\delta$ portion of the payment of all Aliens and Bills, which can be calculated as:
\begin{align*}
& \tilde{\Pi}_i^B(1,\tilde{\boldsymbol{\alpha}}_{-i}) \\
= & T \delta \tilde{p}_i \left(\sum_{j \in \Ku \bigcap \Ka} \eta_{j,i} \tilde{\sigma}_{j,i}^{B\ast}+ \sum_{j \in \Ku \bigcap \Ks,j\neq i} \eta_{j,i} \tilde{\alpha}_j \tilde{\sigma}_{j,i}^{B\ast}\right).
\end{align*}

Based on the above, the perceived payoff of Bill $i$ is
\begin{equation}\label{eq:NewtotalpayoffBill}
\widetilde{V}_i^B(1,\tilde{\boldsymbol{\alpha}}_{-i}) = \tilde{\Pi}_i^B(1,\tilde{\boldsymbol{\alpha}}_{-i}) + T \sum_{k=0}^K \eta_{i, k} \widetilde{V}_{i,k}^B(0,\tilde{\boldsymbol{\alpha}}_{-i}).
\end{equation}

Regarding the existence of the Nash equilibrium, we have the same conclusions in Theorem \ref{lemmaExiMixedNE} as the membership selection game in the Stackelberg model, in Section \ref{sec:membership}.
We can use the same smoothed best response update algorithm in Section \ref{sec:membership} to derive the Nash equilibrium for the approximate membership selection game.

\begin{figure*}[t]
 \centering
\begin{minipage}[t]{0.3 \linewidth}
\centering
\includegraphics[height=1.8 in]{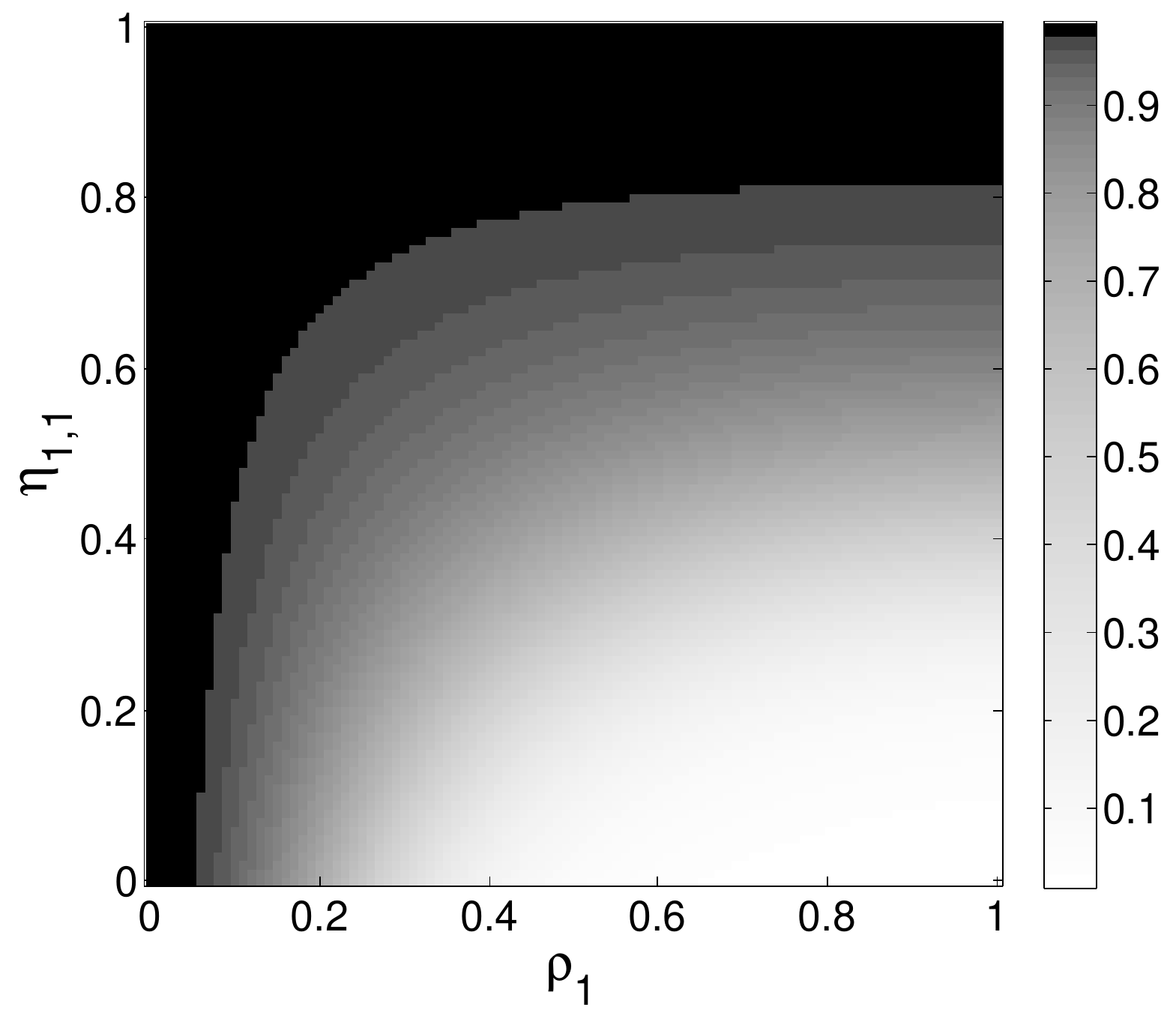}
  \caption{Membership Decision under Different Parameters (Color denoting subscriber $1$'s probability of choosing to be~a~Bill.)}\label{fig:Parameter}
\end{minipage}
\begin{minipage}[t]{0.03 \linewidth}
~
\end{minipage}
\begin{minipage}[t]{0.3 \linewidth}
\centering
\includegraphics[height=1.8 in]{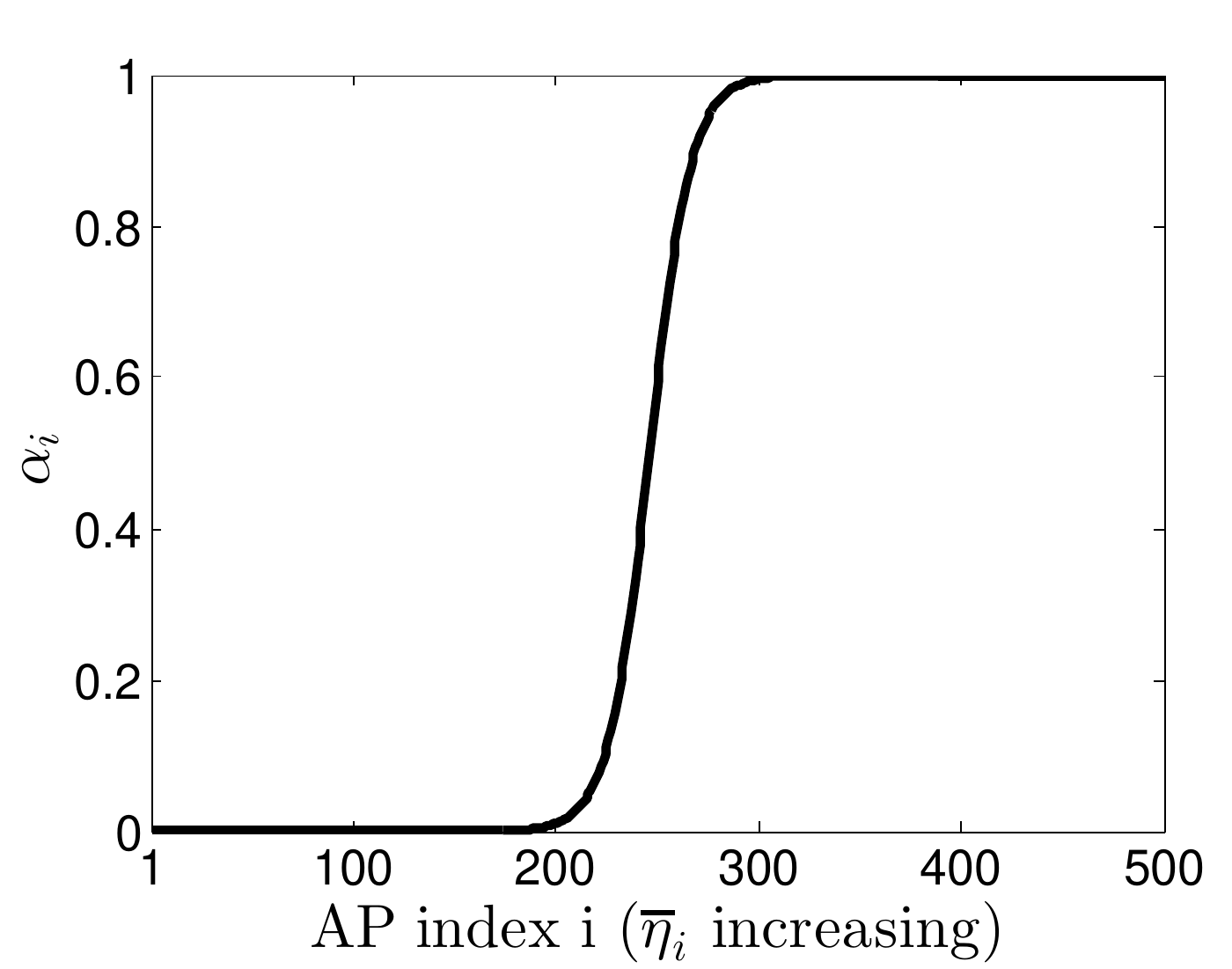}
  \caption{Membership Decisions under Different Location Popularities}\label{fig:LP}
\end{minipage}
\begin{minipage}[t]{0.03 \linewidth}
~
\end{minipage}
\begin{minipage}[t]{0.3 \linewidth}
\centering
\includegraphics[height=1.8 in]{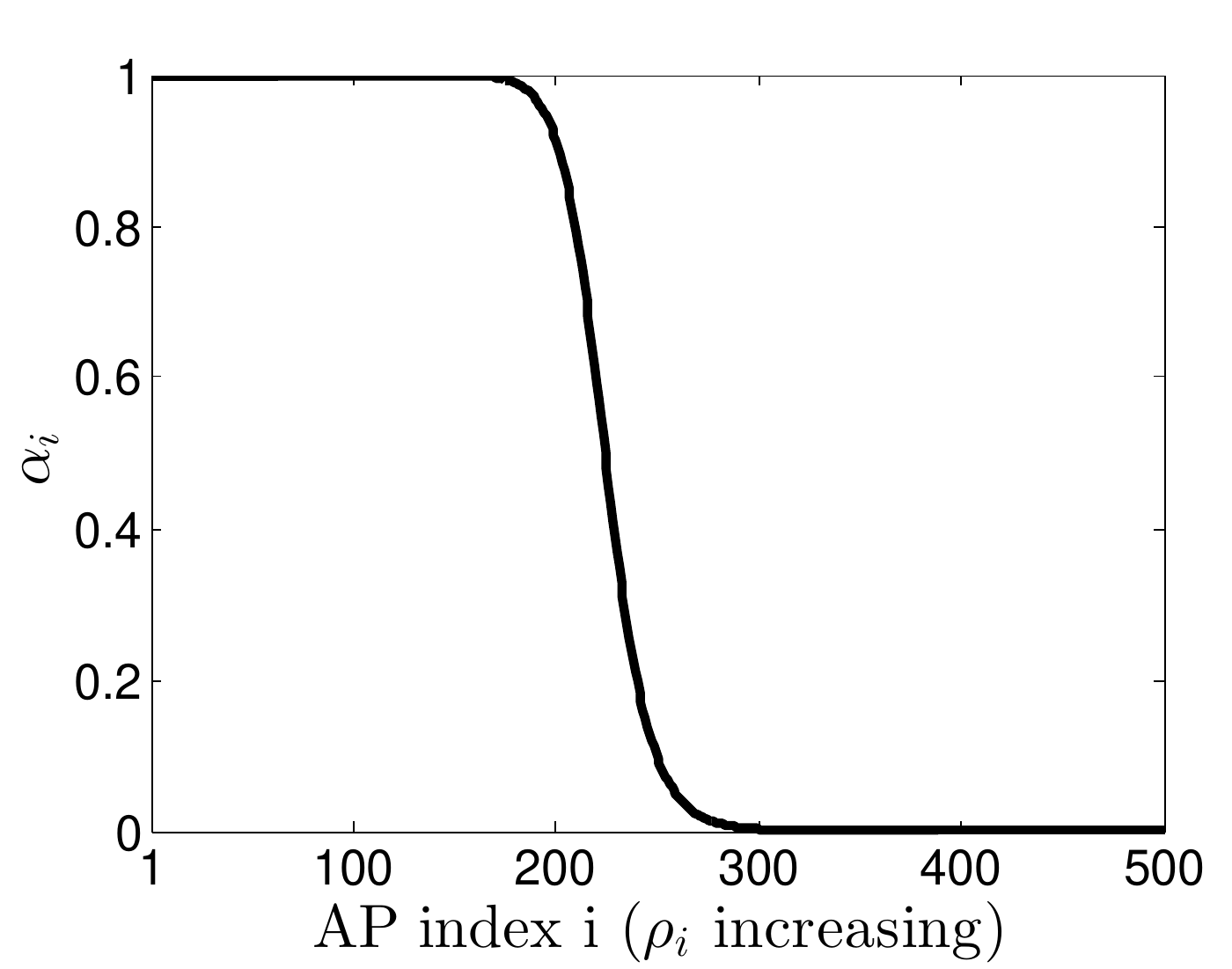}
  \caption{Membership Decisions under Different Network Access Evaluations}\label{fig:UP}
\end{minipage}
\vspace{-3mm}
\end{figure*}

\subsection{Layer I: Operator Pricing Design}

In this section, we study the operator's pricing design in Layer I of the approximate Stackelberg model.
Similarly, we propose a partial price differentiation scheme.

In the partial price differentiation scheme, the operator first segment the APs in the network into different groups, based on AP attributes, as in Section \ref{sec:APsegmentation}.
We denote the set of APs in group $g ~(g=1,2,\cdots,G)$ as set $\widetilde{\mathcal{S}}_g$,
and we denote the AP segmentation result as $\widetilde{\boldsymbol{\mathcal{S}}}=\{\widetilde{\mathcal{S}}_1,\widetilde{\mathcal{S}}_2,\cdots,\widetilde{\mathcal{S}}_G\}$.

Then the operator sets one price to APs in the same group.
The operator's goal is to maximize the total revenue he collected on all APs in the network:
\vspace{-1mm}
\begin{equation*}
\vspace{-2mm}
\widetilde{H}(\tilde{\boldsymbol{p}}^P)=\sum_{i=1}^K \tilde{h}_i(\tilde{\boldsymbol{p}}^P).
\end{equation*}
Here $\tilde{h}_i(\tilde{\boldsymbol{p}}^P)$ is the expected revenue collected by the operator on AP $i$, which can be written as:
\vspace{-1mm}
\begin{equation}
\vspace{-1mm}
\tilde{h}_i(\tilde{\boldsymbol{p}}^P)=\left( 1 \cdot \left( 1-\tilde{\alpha}_i^\ast \right) + \left( 1-\delta \right) \tilde{\alpha}_i^\ast \right) \tilde{p}_g^P \tilde{\sigma}_i^\ast (\tilde{\boldsymbol{p}}^P),
\end{equation}
Here $i\in\widetilde{\mathcal{S}}_g$, and $\tilde{\sigma}_i^\ast (\tilde{\boldsymbol{p}}^P)$ is the expected \emph{charged} network usage from all Bills (except subscriber $i$) and Aliens at equilibrium on AP $i$:
\vspace{-1mm}
\begin{equation*}
\vspace{-1mm}
\tilde{\sigma}_i^\ast (\tilde{\boldsymbol{p}}^P)= T \tilde{p}_g^P \left(\sum_{j \in \Ku \bigcap \Ka} \eta_{j,i} \tilde{\sigma}_{j,i}^{B\ast}+ \sum_{j \in \Ku \bigcap \Ks,j\neq i} \eta_{j,i} \tilde{\alpha}_j \tilde{\sigma}_{j,i}^{B\ast}\right).\label{eq:EXPusaNew}
\end{equation*}

Given the AP segmentation $\widetilde{\boldsymbol{\mathcal{S}}}=\{\widetilde{\mathcal{S}}_1,\widetilde{\mathcal{S}}_2,\cdots,\widetilde{\mathcal{S}}_G\}$, the operator's partial price differentiation problem in the approximate Stackelberg model can be written as follows:
\begin{align}
\vspace{-1mm}
& \mbox{\textbf{Problem 3: Approximate Partial Price Differentiation}} \notag\\
& \displaystyle \max ~~ \displaystyle
\widetilde{H}(\tilde{\boldsymbol{p}}^P)=\sum_{i=1}^K \tilde{h}_i(\tilde{\boldsymbol{p}}^P) \notag\\
& \mbox{ var:} ~~~  \tilde{p}_g^P  \geq 0, \forall g=1,2,\cdots, G \notag
\end{align}

Problem 3 has similar structure as Problem 2.
Hence, we propose to use DYCORS algorithm to solve Problem 3.

\section{Simulation Results}\label{sec:simu}

We provide simulation results to illustrate the users' behaviors in Section \ref{subsec:user} and evaluate the operator's revenue in Section \ref{subsec:operator}.

\subsection{User Behaviors}\label{subsec:user}

Regarding user behaviors, we numerically study how the network access valuation parameter $\rho_i$ and the mobility pattern $\boldsymbol{\eta}_i$ affect subscriber $i$'s membership selection decision, given other system parameters fixed.
In what follows, we will first simulate a small network with $2$ APs (subscribers) and $1$ Alien, to gain insights of a single user's best choice.
Then, we will simulate a large network with $500$ APs and $500$ Aliens to understand the system-level performance.

\subsubsection{A Small Network Example}\label{subsec:small}

We simulate a small network with $2$ subscribers (each owns an AP) and $1$ Alien.
We study how subscriber $1$'s network access valuation parameter $\rho_1$ and his probability of staying at home $\eta_{1,1}$ affect his membership selection.

We assume that the revenue sharing ratio $\delta=0.5$. Both APs have the same price $p=1$.
In Section \ref{subsec:operator}, we will consider the operator's price optimization problem.
The mobility patterns of subscriber $2$ and the Alien are the same: $\boldsymbol{\eta}_2=\boldsymbol{\eta}_a=[1/3,1/3,1/3]$.
We assume that $\rho_1 \in [0,1]$. Subscriber $1$ stays at home with probability $\eta_{1,1}$, and travels to AP $2$ and outside the Wi-Fi coverage with a same probability $\eta_{1,2}=\eta_{1,0}=(1-\eta_{1,1})/2$.

Figure \ref{fig:Parameter} shows subscriber $1$'s membership selection decision in the equilibrium (Definition \ref{NEImixed} in Section \ref{subsec:mexed}), under different values of $\rho_i \in [0,1]$ and $\eta_{1,1} \in [0,1]$.
The color represents the value of $\alpha_1$, which is subscriber $1$'s probability of choosing to be a Bill.
The black region corresponds to $\alpha_1=1$, and the white region corresponds to $\alpha_1=0$.
The color in between corresponds to a mixed strategy of $\alpha_1\in (0,1)$, as shown in the colorbar on the right.

Figure \ref{fig:Parameter} shows that when $\eta_{1,1}$ is large enough (i.e., larger than $0.82$), i.e., subscriber $1$ stays at home most of the time, his will always choose to be a Bill (with the probability $\alpha_1=1$), independent of subscriber $2$'s membership decision.
As $\eta_{1,1}$ becomes smaller and $\rho_1$ becomes larger, the performance and payment during roaming becomes increasingly important, so subscriber $1$ starts to choose a mixed strategy with a smaller number of $\alpha_1$.
When $\eta_{1,1}$ is small enough and $\rho_1$ is large enough, e.g., the right bottom corner of Figure \ref{fig:Parameter}, he will always choose to be a Linus with a probability $1-\alpha_1=1$.

\subsubsection{Simulation Results for Large Network}\label{subsec:large}

Next, we simulate a larger network with $500$ APs and $500$ Aliens.
We assume the price $p=1$ and the revenue sharing ratio $\delta=0.5$ .
We study how the system parameters, i.e., the location popularity and the network access evaluation, affect subscribers' membership decisions.

We first study how the location popularity of an AP affects the subscriber's membership selection decision.
We assume that each subscriber's network access valuation parameter $\rho_i=1,\forall i\in\Ks$.
Since Aliens do not have APs, we assume that Aliens value the network access more, and $\rho_i=5,\forall i \in \Ka.$
We assume that all users show up at a particular AP, e.g., AP $i$ ($i \in \Ks$), with the same probability $\bar{\eta}_i$, i.e.,
%\vspace{-2mm}
\begin{equation*}
%\vspace{-1mm}
\eta_{j,i}=\bar{\eta}_i, \forall j\in \Ku .
\end{equation*}
We denote $\bar{\eta}_0$ as the probability of users showing up at areas that are not covered by any WiFi AP. Obviously, we have:
\vspace{-1mm}
\begin{equation*}
\vspace{-2mm}
\sum_{i=0}^{K}\bar{\eta}_i=1 .
\end{equation*}
For simplicity, we assume that the location popularity $\bar{\eta}_i$ increases with the AP index $i$.

Figure \ref{fig:LP} shows each of the $500$ subscribers' membership selection decision.
Simulation result shows that the subscriber's probability of choosing to be a Bill increases with his AP location popularity.
The reason is that a subscriber whose AP is located at a more popular location can earn more revenue from other Bills and Aliens.

We then study how the network access evaluation of a subscriber affects his membership selection decision.
We assume that the probability of each user showing up at each location is uniform and same.
We assume that the subscribers' network access evaluation parameters $\rho_i$ increases with the AP index $i$.

Figure \ref{fig:UP} shows each of the $500$ subscribers' membership selection decision.
Simulation result shows that the subscriber's probability of choosing to be a Bill decreases with his network access evaluation.
This is because he cares more about the network access benefit when roaming, and hence is more willing to be a Linus to enjoy free access and consume more data during roaming.

\begin{figure}
  \centering
  \includegraphics[width=0.36\textwidth]{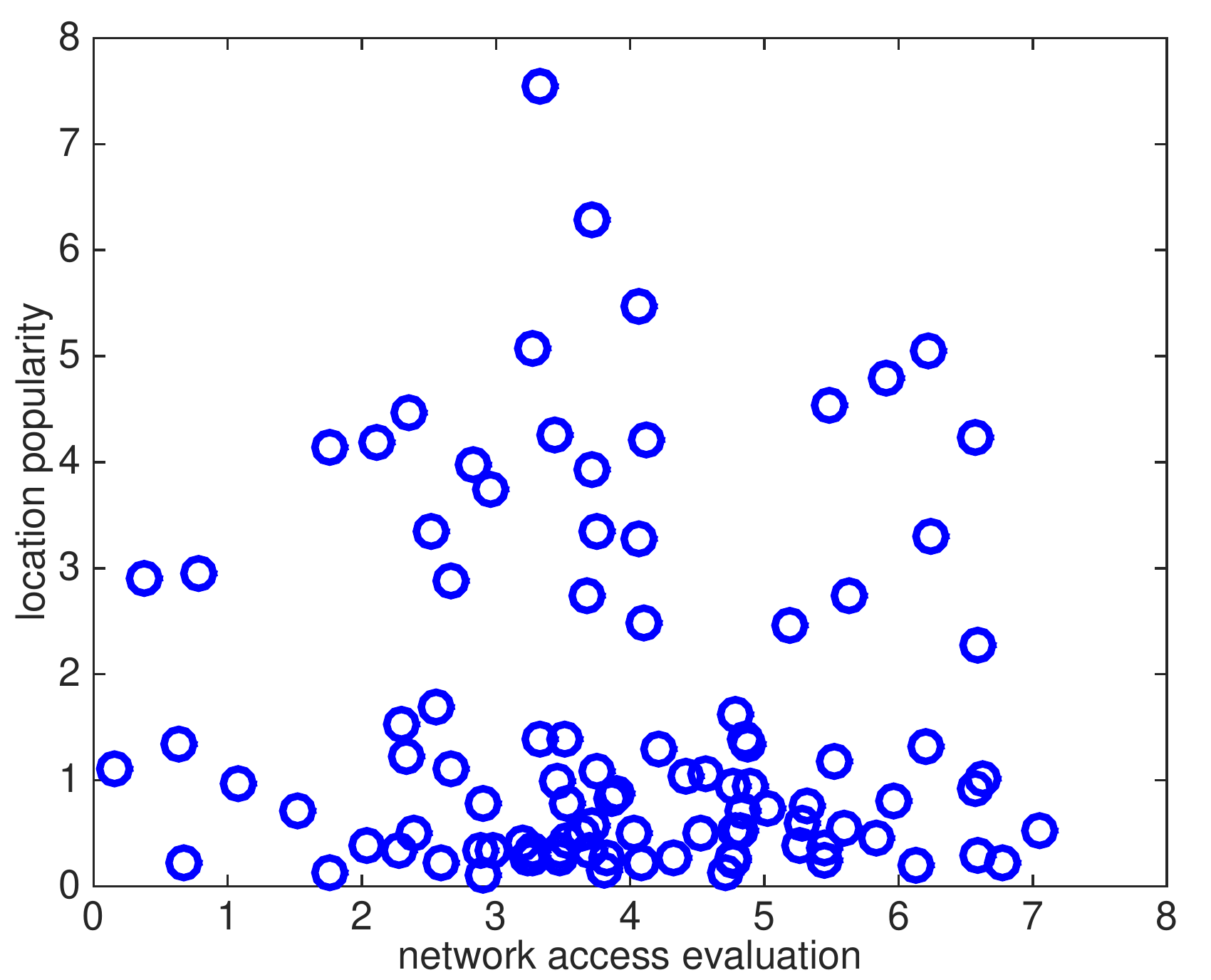}
  \vspace{-3mm}
 \caption{Subscribers' Network Access Evaliation and Location Popularity Parameters}\label{fig:ParameterLarge}
 \vspace{-3mm}
\end{figure}

\subsection{The Operator's Revenue}\label{subsec:operator}

Regarding the operator's revenue, we numerically study the performance of the partial price differentiation scheme in a large network.
We consider a network which consists of 100 APs and 100 Aliens.

We assume that subscribers' network access evaluation parameters $\rho_i,\forall i \in \Ks$ follow the Gaussian distribution with a mean value of $4$ and a variance of $2$.
We obtain the APs' location popularity $\bar{\eta}_i=\sum_{j\neq i}\eta_{j,i} ~(\forall i \in \Ks)$ based on the the realistic mobility data in \cite{mobility}.
Figure \ref{fig:ParameterLarge} shows the relationship between network access evaluation parameter and location popularity parameter.

We study the performance of the partial price differentiation scheme.
We first use the weighted k-means clustering algorithm to segment APs into different groups considering the attributes of network access evaluation and location popularity.
We assign a weight $\beta$ to the attribute of network access evaluation and $1-\beta$ to the attribute of location popularity in the weighted k-means clustering algorithm.
The AP segmentation results under different weight $\beta$ are provided in the appendix,
%online appendix \cite{report}.
Then we apply the price differentiation scheme and optimize over the prices to maximize the operator's revenue.

\begin{figure} 
 \centering
  \includegraphics[width=0.36\textwidth]{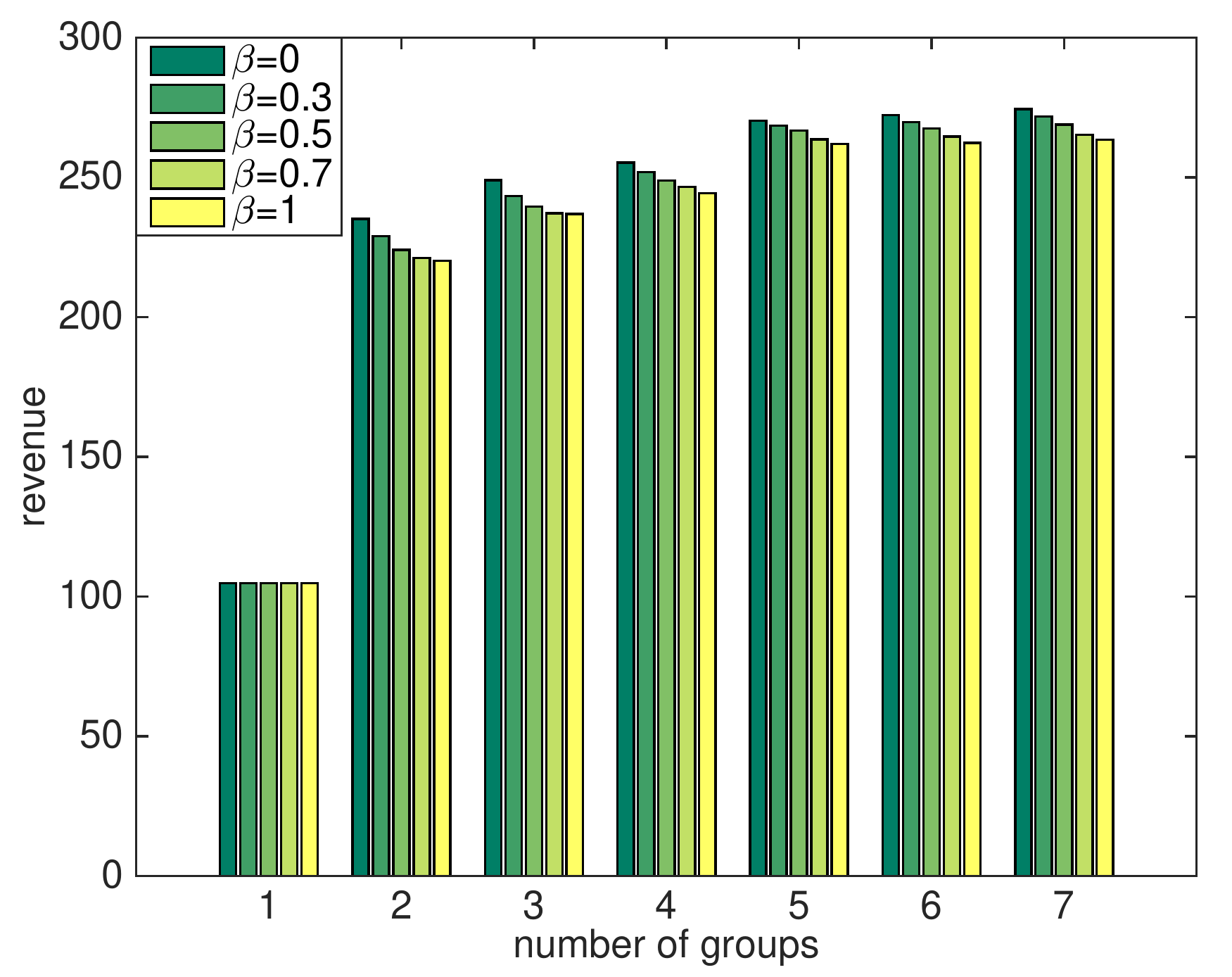}
  \vspace{-3mm}
 \caption{Operator's Revenue under the Patial Price Differentiation}\label{fig:OperatorRevenue}
\vspace{-3mm}
\end{figure}

Figure \ref{fig:OperatorRevenue} shows the operator's revenue under different number of groups and different weight $\beta$.
Given a particular weight $\beta$, we can see that
(i) the operator's revenue increases with the number of groups; and (ii) the revenue increase rate slows down as the number of groups increase.
Given a particular number of groups, the revenue difference under different $\beta$ is small.
Compared with the single pricing scheme, the two-price scheme can effectively increase the operator's revenue by up to $124.44\%$.
However, the revenue increase from the 6-price scheme to the 7-price scheme is only $0.78\%$.
Hence, the operator can achieve a good tradeoff of total revenue and implementation complexity, by applying the partial price differentiation scheme.

\section{Conclusion}\label{sec:conc}

In this paper, we analyze the user behavior and the operator pricing design for the crowdsourced wireless community network.
We model the interactions between the operator and users as a Stackelberg model where users are fully rational in small-scale systems.
For the large-scale systems where users are bounded rational due to the limited computation capability, we propose and analyze an approximate Stackelberg model.
We study the user behaviors under both full rationality and bounded rationality systematically.
We show that a user with a more popular home location, a smaller probability of travelling, or a smaller network access evaluation is more likely to choose to be a Bill.
Moreover, we propose a partial price differentiation scheme for the operator, based on the analysis of users response to the pricing scheme.
Our numerical results show that the proposed partial price differentiation scheme with only two prices can increase the operator's revenue up to 124.44\% comparing with the single pricing scheme, and can achieve an average of 80\% of the maximum operator revenue under the complete price differentiation scheme.

There are several interesting and important extensions for the model in this work.
For example, it is important to study the model under information asymmetry, where users have some private information, e.g., network access valuation $\rho$ and {user mobility pattern $\boldsymbol{\eta}$}.
In that case, the operator needs to further design an incentive compatible mechanism (e.g., auction) to elicit the users' private information.

% biography section
%
% If you have an EPS/PDF photo (graphicx package needed) extra braces are
% needed around the contents of the optional argument to biography to prevent
% the LaTeX parser from getting confused when it sees the complicated
% \includegraphics command within an optional argument. (You could create
% your own custom macro containing the \includegraphics command to make things
% simpler here.)
%\begin{IEEEbiography}[{\includegraphics[width=1in,height=1.25in,clip,keepaspectratio]{mshell}}]{Michael Shell}
% or if you just want to reserve a space for a photo:

\begin{IEEEbiography}[{\includegraphics[width=1in,height=1.25in,clip,keepaspectratio]{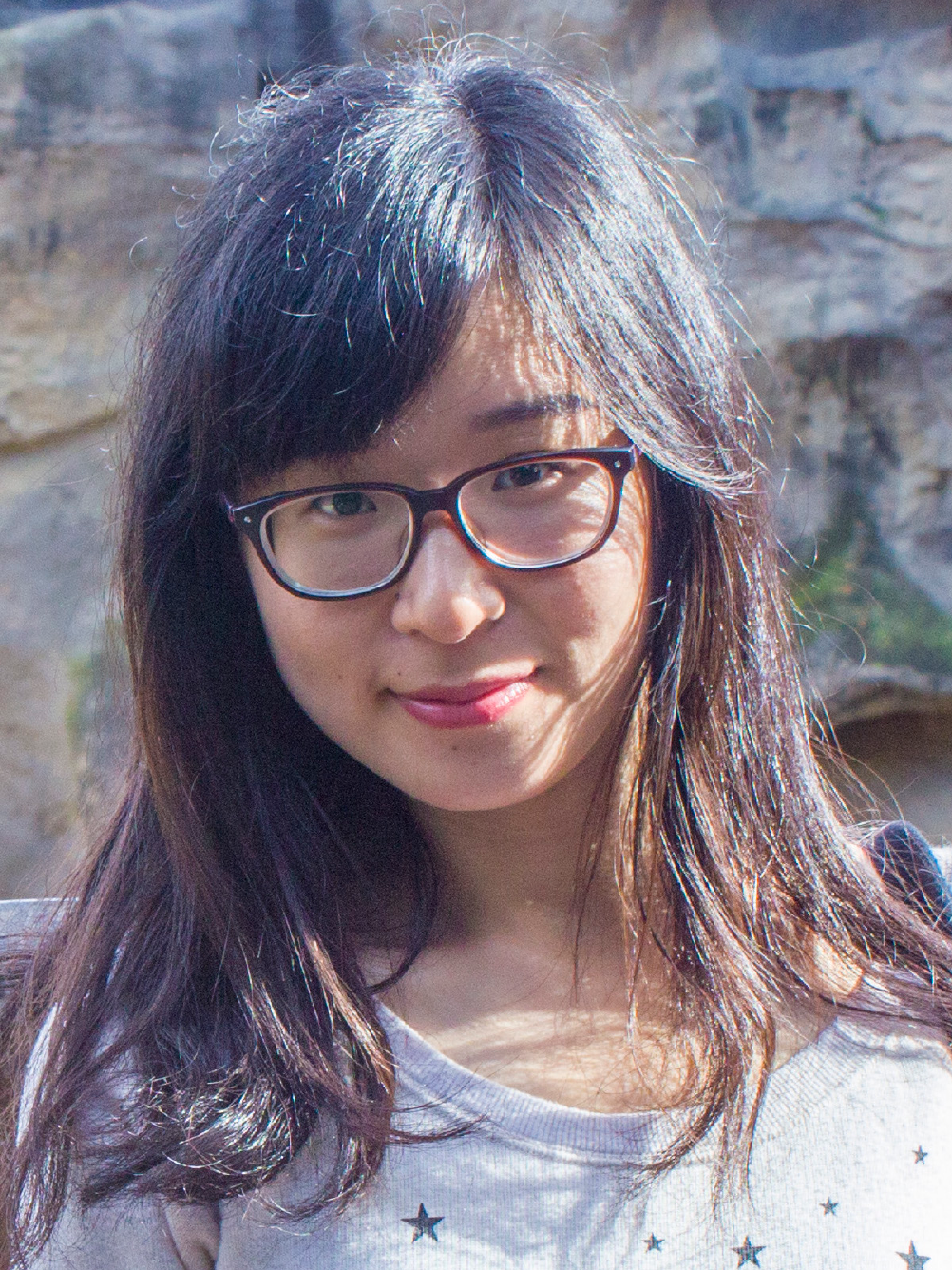}}]{Qian Ma}
(S'13) is a Ph.D. student in the Department of Information Engineering at the Chinese University of Hong Kong.
She received the B.S. degree from Beijing University of Posts and Telecommunications (China) in 2012.
Her research interests lie in the field of wireless communications and network economics.
She is the recipient of the Best Student Paper Award from the IEEE International Symposium on Modeling and Optimization in Mobile, Ad
Hoc and Wireless Networks (WiOpt) in 2015.
\end{IEEEbiography}

\begin{IEEEbiography}[{\includegraphics[width=1in,height=1.25in,clip,keepaspectratio]{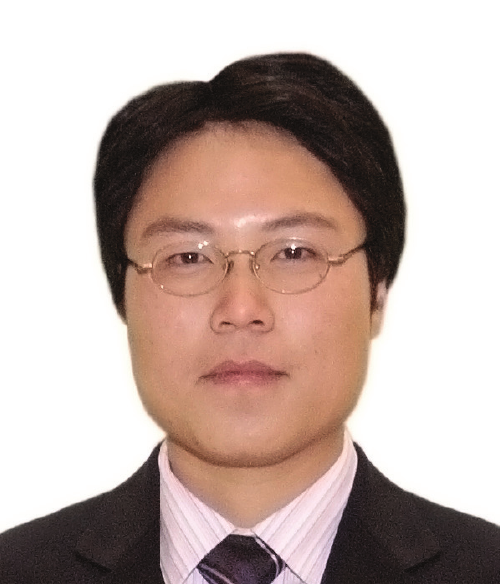}}]{Lin Gao}
(S'08-M'10) is an Associate Professor at Harbin Institute of Technology (Shenzhen), China.
He received M.S. and Ph.D. degrees in Electronic Engineering from Shanghai Jiao Tong University (China), in 2006 and 2010, respectively. He was a Postdoctoral Fellow at The Chinese University of Hong Kong from 2010 to 2015.
His research interests are in the interdisciplinary area combining telecommunications and microeconomics, with particular focus on the game-theoretic and economic analysis for various communication and network scenarios, including cognitive radio networks, TV white space networks, cooperative communications, 5G communications, mobile crowdsensing, and mobile Internet.
\end{IEEEbiography}

\begin{IEEEbiography}[{\includegraphics[width=1in,height=1.25in,clip,keepaspectratio]{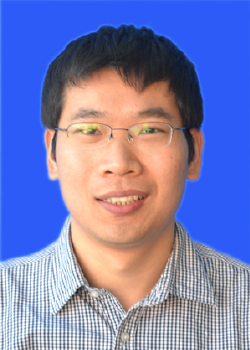}}]{Ya-Feng Liu}
(M'12) is an Assistant Professor of the Academy of Mathematics and Systems Science, Chinese Academy of Sciences, where he received the Ph.D degree in Computational Mathematics in 2012.
His main research interests are nonlinear optimization and its applications to signal processing, wireless communications, and machine learning.
He is a recipient of the Best Paper Award from the IEEE International Conference on Communications (ICC) in 2011 and the Best Student Paper Award from the International Symposium on Modeling and Optimization in Mobile, Ad Hoc and Wireless Networks (WiOpt) in 2015.
\end{IEEEbiography}

\begin{IEEEbiography}[{\includegraphics[width=1in,height=1.25in,clip,keepaspectratio]{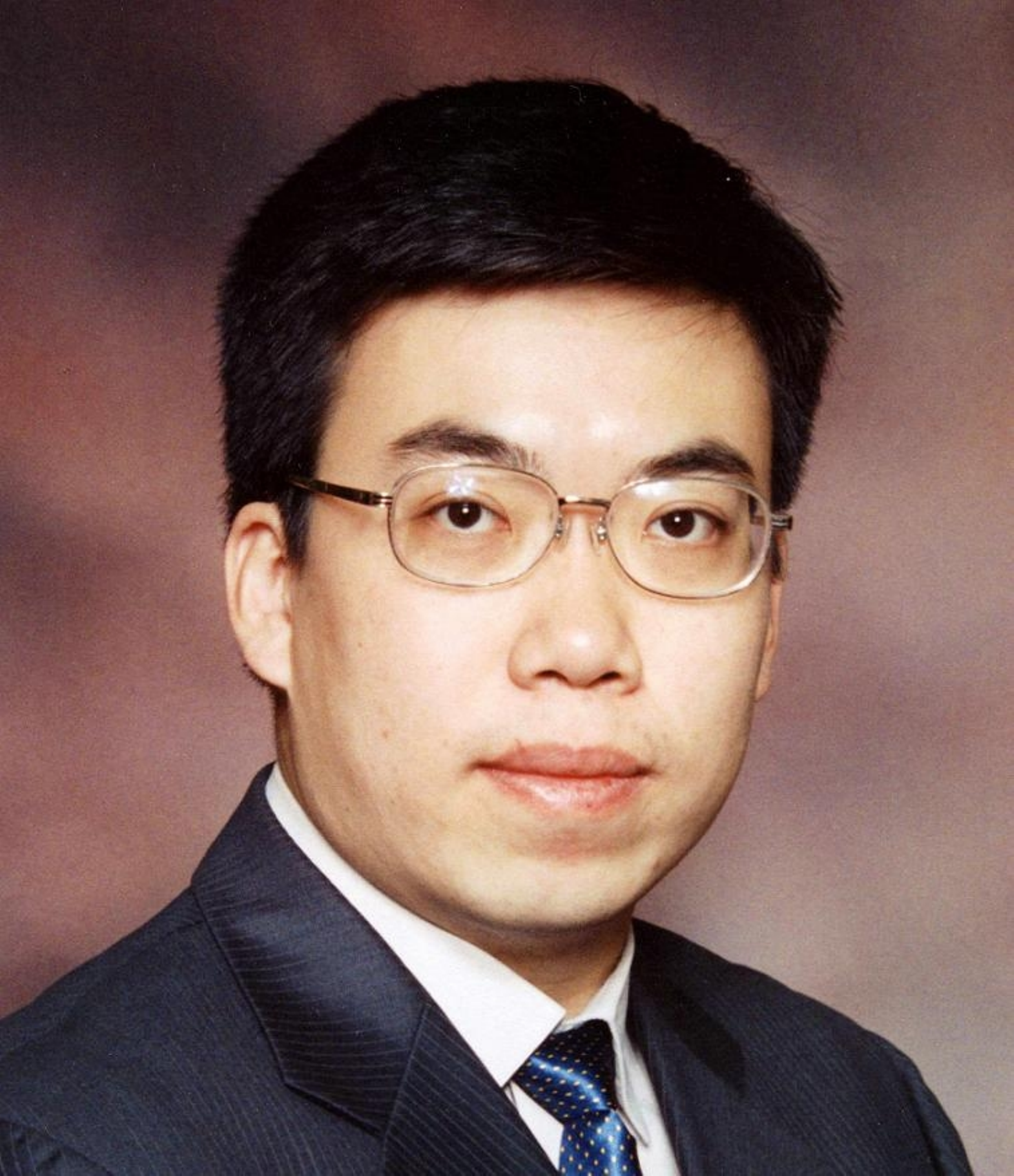}}]{Jianwei Huang}
(S'01-M'06-SM'11-F'16) is an Associate Professor and Director of the Network Communications and Economics Lab (ncel.ie.cuhk.edu.hk), in the Department of Information Engineering at the Chinese University of Hong Kong.
He received the Ph.D. degree from Northwestern University in 2005.
He is the co-recipient of 8 international Best Paper Awards, including IEEE Marconi Prize Paper Award in Wireless Communications in 2011.
He has co-authored five books and six ESI highly cited papers.
He has served as an Editor of several top IEEE Communications journals, including JSAC, TWC, and TCCN.
He is an IEEE Fellow and a Distinguished Lecturer of IEEE Communications Society.
\end{IEEEbiography}

\newpage

%\appendix
\appendices

\section{Proof of Lemma 1} \label{applemmaBRLinus}

\begin{proof}
If user $i$ is a Linus, his payoff on AP $k$ can be derived from Eq. (3), which is an increasing function of $\sigma_{i,k}$.
Hence, regardless of other users' strategies, a Linus-type user $i$'s best response is always to choose the maximum network access time, i.e.,
$\sigma_{i,k}^\ast = 1$.
\end{proof}

\section{Proof of Lemma 2} \label{applemmaBRBill}

\begin{proof}
If user $i$  is a Bill or an Alien, his payoff at AP $k$ can be derived from Eq. (4). Hence, the payoff maximization problem for user $i$ on AP $k$ can be written as follows
\begin{align}
& \displaystyle \max ~~ \rho_i \log(1+\bar{r}_{i,k}(\boldsymbol{\sigma}_{-i,k}) \cdot \sigma_{i,k}) - p \sigma_{i,k} \notag\\
& \mbox{~s.t.} ~~~~ 0 \leq \sigma_{i,k} \leq 1 \notag\\
& \mbox{~var:} ~~~ \sigma_{i,k} \notag
\end{align}
which is a concave maximization problem. By using the first-order optimality condition, we can derive a Bill-type or Alien user $i$'s best response:
\begin{equation*}
\sigma_{i,k}^\ast = \min\left\{ 1,\max\left\{ \frac{\rho_i}{p}-\frac{1}{\bar{r}_{i,k}(\boldsymbol{\sigma}_{-i,k})},0 \right\} \right\},
\end{equation*}
which is a function of other users' strategies $\boldsymbol{\sigma}_{-i,k} $.
\end{proof}

\section{Proof of Theorem 1} \label{applemmaExiNEI}

\begin{proof}
To prove Theorem 1, we first introduce the Brouwer fixed-point theorem [17],
which states that every continuous mapping $f$ from a compact convex set to itself has a fixed point $z_0$ satisfying $z_0=f(z_0).$

The Nash equilibrium $\boldsymbol{\sigma}_k^\ast=\{\sigma_{i,k}^\ast,\forall i \in \K(k)\}$ is computed by Lemma 1 and Lemma 2.
Specifically,
\begin{equation*}
\sigma_{i,k} =\\
\begin{cases}
1, \mbox{ Linus} ~i;
\\
\min\left\{ 1,\max\left\{ \frac{\rho_i}{p}-\frac{1}{\bar{r}_{i,k}(\boldsymbol{\sigma}_{-i,k})},0 \right\} \right\}, \mbox{ Bill or Alien} ~i.
\end{cases}
\end{equation*}
This is a continuous mapping from a compact convex set $[0,1]$ to itself, which satisfies the requirement of the Brouwer fixed-point theorem.
Hence, there exists a fixed point $\boldsymbol{\sigma}_k^\ast$ satisfying
\begin{equation*}
\sigma_{i,k}^\ast =\\
\begin{cases}
1, \mbox{ Linus} ~i;
\\
\min\left\{ 1,\max\left\{ \frac{\rho_i}{p}-\frac{1}{\bar{r}_{i,k}(\boldsymbol{\sigma}_{-i,k}^\ast)},0 \right\} \right\}, \mbox{ Bill or Alien} ~i.
\end{cases}
\end{equation*}
The fixed point $\boldsymbol{\sigma}_k^\ast$ is the Nash equilibrium.
\end{proof}

\section{Proof of Proposition 1}\label{appUniNEI}

Assume there are $2$ APs and $1$ Alien in the network, and we study the two-player network access game on AP $2$. The two players are subscriber $1$ and the Alien, and their strategies are $\sigma_{1,2}\in[0,1]$ and $\sigma_{a,2}\in[0,1]$. We have already shown that there exists a Nash equilibrium $(\sigma_{1,2}^\ast,\sigma_{a,2}^\ast)$ where:
\begin{equation*}
\sigma_{1,2}^\ast =\\
\begin{cases}
1, \mbox{ if } ~x_1=0;
\\
\min\left\{ 1,\max\left\{ \frac{\rho_1}{p}-\frac{1}{\bar{r}_{1,2}(\boldsymbol{\sigma}_{a,2}^\ast)},0 \right\} \right\}, \mbox{ if } ~x_1=1;
\end{cases}
\end{equation*}
and
$$\sigma_{a,2}^\ast = \min\{1,\max\{\frac{\rho_a}{p}-\frac{1}{\bar{r}_{a,2}(\sigma_{1,2}^\ast)},0\}\}.$$
Here the expected data rates of subscriber $1$ and the Alien are
\begin{align}
& \bar{r}_{1,2}(\sigma_{a,2}^\ast)=(1-\sigma_{a,2}^\ast)\bar{R}(1)+\sigma_{a,2}^\ast \bar{R}(2), \notag \\
& \bar{r}_{a,2}(\sigma_{1,2}^\ast)=(1-\sigma_{1,2}^\ast)\bar{R}(1)+\sigma_{1,2}^\ast \bar{R}(2). \notag
\end{align}

Now we show Proposition 1 is true.

\begin{proof}
If subscriber $1$ chooses to be a Linus, i.e., $x_1=0$, then $\sigma_{1,2}^\ast=1$, and
$$\sigma_{a,2}^\ast = \min\{1,\max\{\frac{\rho_a}{p}-\frac{1}{\bar{R}(2)},0\}\}.$$
The Nash equilibrium is unique and we can directly derive it.

If subscriber $1$ chooses to be a Bill, i.e., $x_1=1$, then we consider the following cases:

\textbf{A}: If $\frac{\rho_1}{p}-\frac{1}{\bar{R}(1)}<0$, then $\sigma_{1,2}^\ast=0$ and
$$ \sigma_{a,2}^\ast=\frac{\rho_a}{p}-\frac{1}{\bar{R}(1)}. $$

\textbf{B}: If $\frac{\rho_1}{p}-\frac{1}{\bar{R}(2)}>1$, then $\sigma_{1,2}^\ast=1$ and
$$ \sigma_{a,2}^\ast=\frac{\rho_a}{p}-\frac{1}{\bar{R}(2)}. $$

\textbf{C}: If $\frac{\rho_1}{p}-\frac{1}{\bar{R}(1)}\geq 0$ and $\frac{\rho_1}{p}-\frac{1}{\bar{R}(2)}\leq 1$, then we further consider the following three subcases:
\begin{itemize}
\item If $\frac{\rho_a}{p}-\frac{1}{\bar{R}(1)}<0$, then $\sigma_{a,2}^\ast=0$ and
$$ \sigma_{1,2}^\ast=\frac{\rho_1}{p}-\frac{1}{\bar{R}(1)}. $$
\item If $\frac{\rho_a}{p}-\frac{1}{\bar{R}(2)}>1$, then $\sigma_{a,2}^\ast=1$ and
$$ \sigma_{1,2}^\ast=\frac{\rho_1}{p}-\frac{1}{\bar{R}(2)}. $$
\item If $\frac{\rho_a}{p}-\frac{1}{\bar{R}(1)}\geq 0$ and $\frac{\rho_a}{p}-\frac{1}{\bar{R}(2)}\leq 1$, then we have
\begin{align}
& \sigma_{1,2}^\ast = \frac{\rho_1}{p}-\frac{1}{\bar{r}_{1,2}(\sigma_{a,2}^\ast)}, \notag \\
& \sigma_{a,2}^\ast = \frac{\rho_a}{p}-\frac{1}{\bar{r}_{a,2}(\sigma_{1,2}^\ast)}, \notag
\end{align}
which can be denoted as the following mapping
$$ \boldsymbol{\sigma}_2^\ast=T(\boldsymbol{\sigma}_2^\ast) .$$
\end{itemize}

We have:
\begin{align}
& \left|\left(\frac{\rho_1}{p}-\frac{1}{\bar{r}_{1,2}(\sigma_{a,2}^\ast)}\right)-\left(\frac{\rho_1}{p}-\frac{1}{\bar{r}_{1,2}(\sigma_{a,2}^\dag)}\right)\right| \notag \\
=& \left| \frac{1}{\bar{r}_{1,2}(\sigma_{a,2}^\dag)} - \frac{1}{\bar{r}_{1,2}(\sigma_{a,2}^\ast)} \right| \notag \\
=& \left| \frac{1}{R(1)-\sigma_{a,2}^\dag (R(1)-R(2))} - \frac{1}{R(1)-\sigma_{a,2}^\ast (R(1)-R(2))} \right| \notag \\
=& \frac{[R(1)-R(2)]\cdot |\sigma_{a,2}^\ast-\sigma_{a,2}^\dag|}{[R(1)-\sigma_{a,2}^\ast(R(1)-R(2))][R(1)-\sigma_{a,2}^\dag(R(1)-R(2))]}  \notag \\
\leq & c \cdot |\sigma_{a,2}^\ast-\sigma_{a,2}^\dag|, \notag
\end{align}
and similarly
$$ \left|\left(\frac{\rho_a}{p}-\frac{1}{\bar{r}_{a,2}(\sigma_{1,2}^\ast)}\right)-\left(\frac{\rho_a}{p}-\frac{1}{\bar{r}_{a,2}(\sigma_{1,2}^\dag)}\right)\right| \leq c |\sigma_{1,2}^\ast-\sigma_{1,2}^\dag|.$$

The above shows that
$$\| T(\boldsymbol{\sigma}_2^\ast)-T(\boldsymbol{\sigma}_2^\dag) \| \leq c \| \boldsymbol{\sigma}_2^\ast-\boldsymbol{\sigma}_2^\dag \|.$$
The condition in Proposition 1, i.e., $c<1$, implies that the mapping function $T(\cdot)$ is contractive.

The contract mapping implies that the Nash equilibrium $(\sigma_{1,2}^\ast,\sigma_{a,2}^\ast)$ is unique.
\end{proof}

\section{Cases with More Than Two Players}

For the cases with more than two players, the uniqueness of the Nash equilibrium depends on the system parameter in a more complicated fashion. For example, for the case with three players, the Nash equilibrium is unique as long as $2 \max \{\bar{R}_1-\bar{R}_2,\bar{R}_2-\bar{R}_3\}/\min\{\bar{R}_1,\bar{R}_2,2\bar{R}_2-\bar{R}_1,\bar{R}_1+\bar{R}_3-2\bar{R}_2\} <1$. The proof idea is similar as the the proof for the case with two players.

\section{A Best Response Update Algorithm for the Network Access Game}\label{appAlg1}

We further design a Best Response Update Algorithm to derive the Nash equilibrium.
The basic idea is as follows.
Given the strategy profile $\boldsymbol{\sigma}_k^n$ at the $n$-th round,
each player computes the corresponding best response.
We denote the best responses of all players by $T(\boldsymbol{\sigma}_k^n)$.
Each user updates the strategy at the $(n+1)$-th round as the best response in $T(\boldsymbol{\sigma}_k^n)$.

\begin{algorithm}[h]
\caption{Best Response Update Algorithm}
\label{algo:br1}
\begin{algorithmic}[1]
\REQUIRE
$\boldsymbol{\sigma}_k^0,\varepsilon . $
\ENSURE
$ \boldsymbol{\sigma}_k^\ast . $
\STATE Set $n = 0$ and $Flag = 0$.
\WHILE {$Flag = 0$}
\STATE Calculate $\boldsymbol{\sigma}_k^{n+1}= T(\boldsymbol{\sigma}_k^n)  $.
\IF {$|\boldsymbol{\sigma}_k^{n+1}-\boldsymbol{\sigma}_k^n|\leq \varepsilon$}
\STATE Set $Flag = 1$.
\ENDIF
\STATE Set $n = n+1$.
\ENDWHILE
\STATE Set $ \boldsymbol{\sigma}_k^\ast=\boldsymbol{\sigma}_k^n. $
\end{algorithmic}
\end{algorithm}

As shown in Appendix \ref{appUniNEI}, the mapping $T(\cdot)$ under the condition $c<1$ is contractive in the Network Access Game with two players. This immediately implies that the sequence generated by the Best Response Update Algorithm converges to the unique Nash equilibrium and the convergence rate is linear.

\section{Proof of Lemma 3}\label{applemmaMS}

\begin{proof}
We first prove the necessity.
If $f_i(\boldsymbol{x}_{-i}^\ast) < 0$, the best membership for subscriber $i$ is $x_i^\ast = 0$ (Linus), hence $(2x_i^\ast-1)\cdot f_i(\boldsymbol{x}_{-i}^\ast) = - f_i(\boldsymbol{x}_{-i}^\ast)> 0 $.
If $f_i(\boldsymbol{x}_{-i}^\ast) > 0$, the best membership for subscriber $i$ is $x_i^\ast = 1$ (Bill), hence $(2x_i^\ast-1)\cdot f_i(\boldsymbol{x}_{-i}^\ast) =  f_i(\boldsymbol{x}_{-i}^\ast)> 0 $.
Then we prove the sufficiency.
Suppose $f_i(\boldsymbol{x}_{-i}^\ast) < 0$. Then the desired $x_i^*$ that satisfies the condition in Lemma 3 is $x_i^* = 0$, which is obviously the best strategy of subscriber $i$.
Suppose $f_i(\boldsymbol{x}_{-i}^\ast) > 0$. Then the desired $x_i^*$ that satisfies the condition in Lemma 3 is $x_i^* = 1$, which is also the best strategy of subscriber $i$.
\end{proof}

\section{Proof of Proposition 2}\label{applemmaETA}

\begin{proof}
By (10),  we notice that a subscriber $i$'s will choose to be a Bill (i.e., $x_i = 1$) if $V_i ( 1,\boldsymbol{x}^\ast_{-i}) > V_i (  0,\boldsymbol{x}^\ast_{-i})$, i.e., 
\begin{equation*}
\begin{aligned}
\delta \cdot \bar{\Pi}_i  (\boldsymbol{x}^\ast_{-i})
& >
\sum_{k\in \Ks}
\eta_{i, k} \cdot
\left(
V_{i,k}(0,\boldsymbol{x}^\ast_{-i}) 
-
V_{i,k}(1,\boldsymbol{x}^\ast_{-i}) 
\right)
\\
& = 
\sum_{k\in \Ks/\{i\}}
\eta_{i, k} \cdot
\left(
V_{i,k}(0,\boldsymbol{x}^\ast_{-i}) 
-
V_{i,k}(1,\boldsymbol{x}^\ast_{-i}) 
\right).
\end{aligned}
\end{equation*}
The equality in the second line follows because $V_{i,i}(0,\boldsymbol{x}^\ast_{-i})  = V_{i,i}(1,\boldsymbol{x}^\ast_{-i})$. 
This is because a subscriber does not need to pay for using his own AP, either as a Bill or as a Linus.  Hence, he will achieve the same payoff on his own AP regardless of his membership selection. 

We further notice that if 
$$
\eta_{i,i} > \underline{\eta}_{i} \triangleq
1 - \frac{\delta \cdot \bar{\Pi}_i  (\boldsymbol{x}^\ast_{-i})}{
\sum_{k\in \Ks/\{i\}} \left(V_{i,k}(0,\boldsymbol{x}^\ast_{-i}) - V_{i,k}(1,\boldsymbol{x}^\ast_{-i})\right)} ,
$$
then, we have: 
$$
\delta \cdot \bar{\Pi}_i  (\boldsymbol{x}^\ast_{-i}) 
> 
(1 - \eta_{i,i}) \cdot \sum_{k\in \Ks/\{i\}} \left( V_{i,k}(0,\boldsymbol{x}^\ast_{-i}) - V_{i,k}(1,\boldsymbol{x}^\ast_{-i})\right).
$$
It is easy to see that the right-hand side of the above inequality is larger than 
$$
\sum_{k\in \Ks/\{i\}}
\eta_{i, k} \cdot
\left(
V_{i,k}(0,\boldsymbol{x}^\ast_{-i}) 
-
V_{i,k}(1,\boldsymbol{x}^\ast_{-i}) 
\right)
$$
as $1 - \eta_{i,i} = \sum_{j \neq i}\eta_{i,j} \geq \eta_{i,k} $, $\forall k \in \Ks/\{i\}$. 
This implies that if the condition $\eta_{i,i} > \underline{\eta}_{i} $ holds, then $V_i ( 1,\boldsymbol{x}^\ast_{-i}) > V_i (  0,\boldsymbol{x}^\ast_{-i})$, and accordingly, subscriber $i$'s best strategy is to choose Bill.

\end{proof}

\section{A Simple Example Showing Inexistence of a Pure Strategy Nash Equilibrium for the Membership Selection Game}\label{appPSNE}

Here we provide a small example with 3 subscribers \{1, 2, 3\}, where the pure-strategy Nash equilibrium does not exist. 
Consider the following example: 
\begin{itemize}
\item Subscriber 1 travels to APs 2 and 3 with the same probability 0.3 and stays at home with the probability 0.4, i.e.,  $\boldsymbol{\eta}_1 = (0, 0.4, 0.3, 0.3)$; 
\item Subscriber 2  only travels to AP 3 (i.e., never travels to AP 1) with the probability 0.3 and stays at home with the probability 0.7, i.e.,  $\boldsymbol{\eta}_1 = (0, 0.7, 0, 0.3)$;  
\item
Subscriber 3 always stays at home, i.e., $\boldsymbol{\eta}_3 = (0, 1, 0, 0)$; 
\end{itemize}

Obviously, subscriber 3's best strategy is always Bill, regardless of the membership selections of other two subscribers. 

Next we observe the best strategies of subscribers 1 and 2. For illustrative purpose, we assume the following payoff or revenue for subscriber 1:  
\begin{itemize}
\item
The total expected payment on AP 1 (from Aliens) is 1.0, with a portion $\delta = 12\% $ is transferred to subscriber 1 when choosing to be a Bill;
\item
The expected payoffs on AP 2 are 1.5 when choosing Linus, and 1.2 when choosing Bill; 
\item 
The expected payoffs on AP 3 are 1.0 when choosing Linus, and 0.8 when choosing Bill, supposing that subscriber 2 is a Bill; 
\item 
The expected payoffs on AP 3 are 0.7 when choosing Linus, and 0.6 when choosing Bill, supposing that subscriber 2 is a Linus; 
\end{itemize}

It is easy to see that if subscriber 2 is a Bill, the best choice of subscriber 1 is Linus, because  
$$
1.5\times 0.3+1.0\times 0.3 > 1.2 \times 0.3+  0.8\times 0.3 + 1.0 \times 12\%, 
$$
while if subscriber 2 is a Linus, the best choice of subscriber 1 is Linus, because 
$$
1.5\times 0.3 + 0.7 \times 0.3 < 1.2 \times 0.3+ 0.6\times 0.3 + 1.0 \times 12\%.
$$ 

Furthermore, we assume the following payoff or revenue for subscriber 2: 
\begin{itemize}
\item
The total expected payment on AP 2 (from subscriber 1 and Aliens) is 1.0 (supposing that subscriber 1 is a Bill), with a portion $\delta = 12\% $ is transferred to subscriber 2 when choosing to be a Bill. 
\item
The total expected payment on AP 2 (from Aliens) is 0.2 (supposing that subscriber 1 is a Linus), with a portion $\delta = 12\% $ is transferred to subscriber 2 when choosing to be a Bill. 
\item 
The expected payoffs on AP 3 are 1.0 when choosing Linus, and 0.8 when choosing Bill, supposing that subscriber 1 is a Bill; 
\item 
The expected payoffs on AP 3 are 0.7 when choosing Linus, and 0.6 when choosing Bill, supposing that subscriber 1 is a Linus; 
\end{itemize}

It is easy to see that if subscriber 1 is a Bill, the best choice of subscriber 2 is Bill, because  
$$
1.0\times 0.3 < 0.8\times 0.3 + 1.0 \times 12\%, 
$$
while if subscriber 1 is a Linus, the best choice of subscriber 1 is Linus, because 
$$
0.7 \times 0.3 > 0.6 \times 0.3 + 0.2 \times 12\%.
$$ 

Obviously, in the above example, subscriber 1 seeks to choose the different membership  as subscriber 2, while subscriber 2 seeks to choose the same membership as subscriber 1. Hence, it is easy to check that there is no pure-strategy Nash equilibrium.

\section{Proof of Theorem 2}\label{appMNE}

\begin{proof}
The Membership Selection Game is a finite game, with $K$ players and each player having two strategies (i.e., $0$ and $1$).
Hence, there exists at least one mixed strategy equilibrium (which includes the pure strategy equilibrium as a special case) [18].
\end{proof}

\section{DYCORS Algorithm for the Operator's Price Optimization Problem}

In the DYCORS algorithm, we denote 
$n_0$ as the number of space-filling design points, 
$n$ as the number of previously evaluated points, 
$\mathcal{A}_n=\{\boldsymbol{p}(1),\ldots ,\boldsymbol{p}(n)\}$ as the set of previously evaluated points, 
and $s_n(\boldsymbol{p})$ as the response surface model built using the points in $\mathcal{A}_n$. 
We denote $Nf_{max}$ as the maximum number of function evaluations allowed, and a strict decreasing function $\Upsilon(n)$ as the probability of perturbing a coordinate whose values are in $[0,1]$. 
Detailed discussions regarding the physical meanings of these parameters can be found in \cite{DYCORS}.

\begin{algorithm}[h]
\caption{DYCORS Algorithm \cite{DYCORS}}
\label{algo:DYCORS}
\begin{algorithmic}[1]
\REQUIRE
$K,K_A,\boldsymbol{\eta},\boldsymbol{\rho},\delta.$
\ENSURE
$ \mathbf{p}^\ast,\boldsymbol{\alpha}^\ast. $
\STATE Evaluate the initial points $\mathcal{I}=\{\boldsymbol{p}(1),\ldots ,\boldsymbol{p}(n_0)\}$.\\
\STATE Find the best point found so far $\boldsymbol{p}^{\ast}$. 
\STATE Set $n=n_0,\mathcal{A}_n=\mathcal{I}$.\\
\WHILE{$n < Nf_{max}$}
\STATE Fit/update a response surface model $s_n(\boldsymbol{p})$ using the data points: $\mathcal{B}_n=\{(\boldsymbol{p},H(\boldsymbol{p})):\boldsymbol{p}\in \mathcal{A}_n\}$.\\
\STATE Determine the probability: $\Upsilon(n)$.\\
\STATE Generate trial points $\Omega_n=\{y_{n,1},\ldots ,y_{n,m}\}$ by:\\
\STATE $\mbox{ }~$(1) Select the coordinates to perturb.\\
\STATE $\mbox{ }~$(2) Randomly generate the trial points according to the normal distribution.\\
\STATE $\mbox{ }~$(3) Project the trial points onto the domain $\{p_i:0\leq  p_i\leq \bar{p}, \forall i\in\Ks\}$ (if necessary).\\
\STATE Select the next iterate $\boldsymbol{p}(n+1)$ from $\Omega_n$ that maximizes $s_n(\boldsymbol{p})$.\\
\STATE Compute $\boldsymbol{\alpha}^\ast(\boldsymbol{p}(n+1))$ by the smoothed best response update algorithm, compute $\bar{\sigma}_i^\ast(\boldsymbol{p}(n+1))$ by \eqref{eq:EXPusa}, and compute $H(\boldsymbol{p}(n+1))$.\\
\STATE If $H(\boldsymbol{p}(n+1))>H(\boldsymbol{p}^{\ast})$, then $\boldsymbol{p}^{\ast}=\boldsymbol{p}(n+1)$.\\
\STATE Set $\mathcal{A}_{n+1}=\mathcal{A}_n \cup \{\boldsymbol{p}(n+1)\}$, and reset $n=n+1$.
\ENDWHILE
\STATE Compute $\boldsymbol{\alpha}^\ast$ by the smoothed best response update algorithm.
\end{algorithmic}
\end{algorithm}

\section{Weighted k-Means Algorithm for AP Segmentation}

The AP segmentation problem deals with the set $\Ks=\{1,2,\cdots,K\}$ of APs in the network. 
We use an attribute vector $\boldsymbol{y}_i$ to represent AP $i$. 
If we consider $M$ attributes of APs, then $\boldsymbol{y}_i$ is a $M$-dimension vector. 
Let $\boldsymbol{w}=\{w_1,w_2,\cdots,w_M\}$ be the weights for the $M$ attributes. 
Obviously, $w_m \in [0,1], \forall m=1,2,\cdots,M$, and $\sum_{m=1}^M w_m =1$.

The weighted k-means clustering \cite{weightkMeans} searches for a segmentation of the $K$ APs into $G$ groups, 
i.e., $\boldsymbol{\mathcal{S}}=\{\mathcal{S}_1,\mathcal{S}_2,\cdots,\mathcal{S}_G\}$, 
which is:
$$
\arg \min_{\boldsymbol{\mathcal{S}}} \sum_{g=1}^G \sum_{\boldsymbol{y}_i \in \mathcal{S}_g} \sum_{m=1}^M w_m || y_{i,m}-z_{g,m} ||^2 .
$$
Here $\boldsymbol{z}_g=\{z_{g,1},z_{g,2},\cdots,z_{g,m}\}$ represents the the centroids of group $g$. 

The AP segmentation problem is computationally difficult (NP hard). 
However, there are efficient heuristic algorithms that are commonly employed and converge quickly to a local optimum. 
Detailed algorithm can be referred to \cite{weightkMeans}.

\section{AP Segmentation Results under Different Weights}

\begin{figure}
 %\vspace{-3mm}
 \centering
  \includegraphics[width=0.36\textwidth]{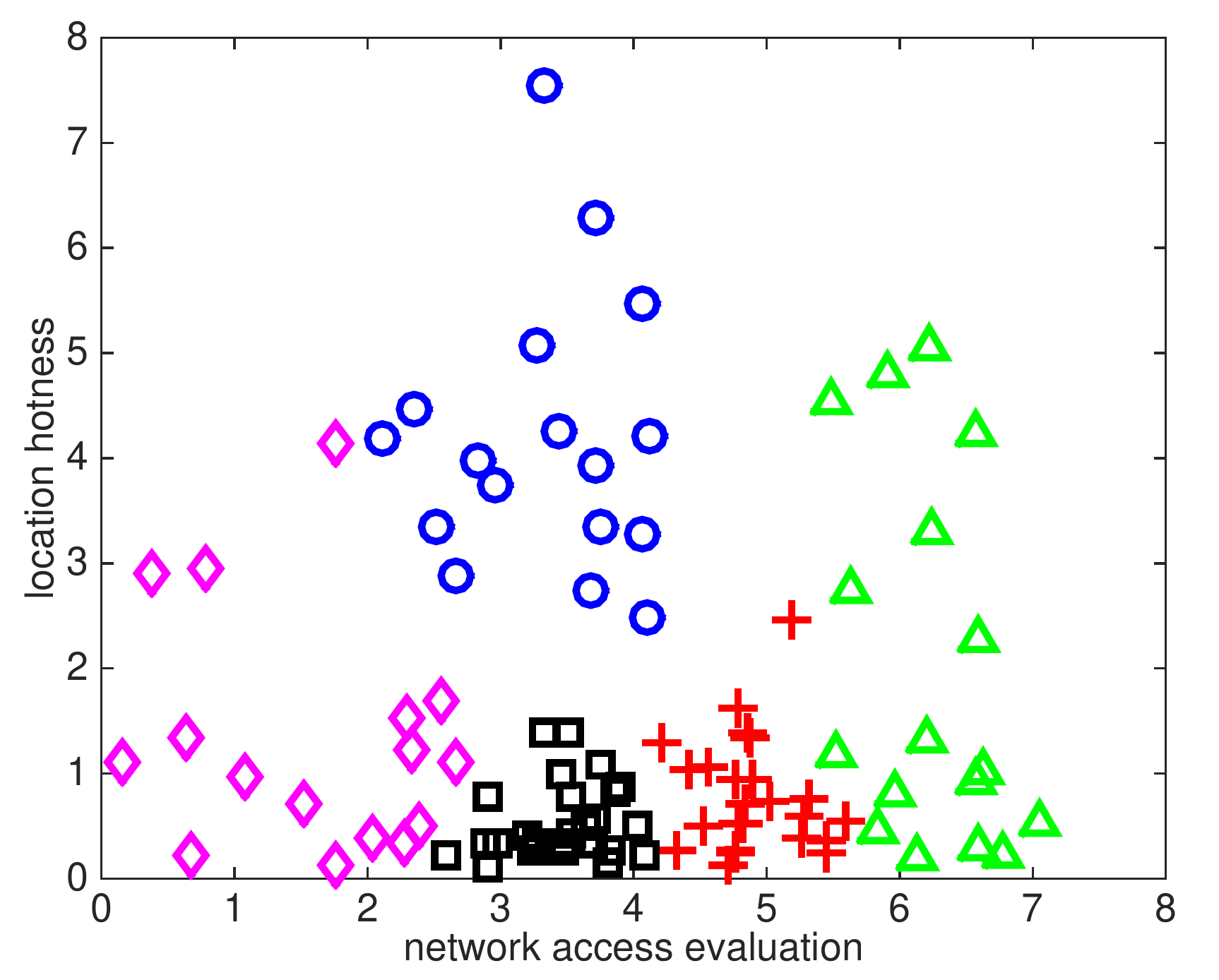}
  \caption{AP Segmentation Result (Five Groups) under $\beta=0.7$}\label{fig:b07g5}
 %\vspace{-3mm}
\end{figure}

\begin{figure}
 %\vspace{-3mm}
 \centering
  \includegraphics[width=0.36\textwidth]{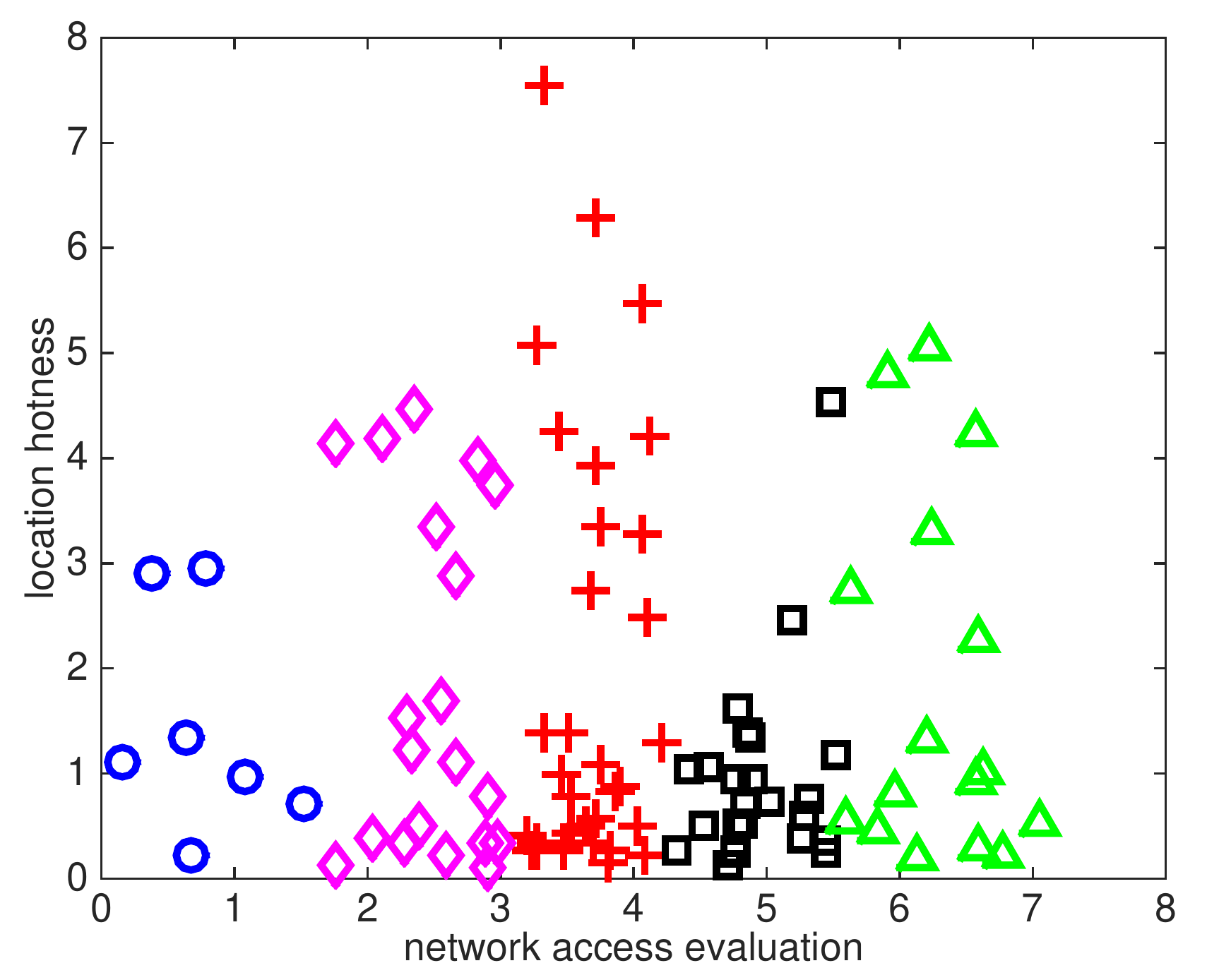}
  \caption{AP Segmentation Result (Five Groups) under $\beta=1$}\label{fig:b1g5}
 %\vspace{-3mm}
\end{figure}

We do the AP segmentation under different weights. 
Specifically, we segment APs under $\beta=0, \beta=0.3, \beta=0.5, \beta=0.7$, and $\beta=1$. 
Recall that $\beta$ is the weight assigned to the network access evaluation parameter, and $1-\beta$ is the weight assigned to the location hotness parameter. 
AP segmentations under different $\beta$ are different. 
Figure \ref{fig:b07g5} shows the AP segmentation result under $\beta=0.7$ where APs are segmented into five groups. 
Figure \ref{fig:b1g5} shows the AP segmentation result under $\beta=1$ where APs are segmented into five groups. 
We can see that when $\beta=1$, the AP segmenation only depends on the network access evaluation parameter. 
However, when $\beta=0.7$, the AP segmentation depends on both the network access evaluation parameter and the location hotness parameter.

\end{document}